\documentclass[iop]{emulateapj}

\usepackage{comment}
\usepackage{natbib}
\usepackage{graphicx}
\usepackage{graphics}
\usepackage{gensymb}
\usepackage{color}
\usepackage{url}
\usepackage{hyperref}
\usepackage{booktabs}
\usepackage{float}
\usepackage{color}
\usepackage{subfigure}
\usepackage{paralist}

\shorttitle{SPIRITS\,15c and SPIRITS\,14buu in IC~2163}
\shortauthors{Jencson et al.}

\begin{document}

\title{SPIRITS\,15c and SPIRITS\,14buu: Two Obscured Supernovae in the Nearby Star-Forming Galaxy IC\,2163}

\author{Jacob E. Jencson\altaffilmark{1,13}, Mansi M. Kasliwal\altaffilmark{1}, Joel Johansson\altaffilmark{2}, Carlos Contreras\altaffilmark{3}, Sergio Castell\'{o}n\altaffilmark{3}, Howard E. Bond\altaffilmark{4,5}, Andrew J. Monson\altaffilmark{4}, Frank J. Masci\altaffilmark{6}, Ann Marie Cody\altaffilmark{7}, Jennifer E. Andrews\altaffilmark{8}, John Bally\altaffilmark{9}, Yi Cao\altaffilmark{1}, Ori D. Fox\altaffilmark{5}, Timothy Gburek\altaffilmark{10}, Robert D. Gehrz\altaffilmark{10}, Wayne Green\altaffilmark{9}, George Helou\altaffilmark{6}, Eric Hsiao\altaffilmark{11}, Nidia Morrell\altaffilmark{3}, Mark Phillips\altaffilmark{3}, Thomas A. Prince\altaffilmark{1}, Robert A. Simcoe\altaffilmark{12}, Nathan Smith\altaffilmark{8}, Samaporn Tinyanont\altaffilmark{1}, and Robert Williams\altaffilmark{5}}

\affil{\altaffilmark{1}Cahill Center for Astrophysics, California Institute of Technology, Pasadena, CA 91125, USA}
\affil{\altaffilmark{2}Benoziyo Center for Astrophysics, Weizmann Institute of Science, 76100 Rehovot, Israel}
\affil{\altaffilmark{3}Las Campanas Observatory, Carnegie Observatories, Casilla 601, La Serena, Chile}
\affil{\altaffilmark{4}Dept. of Astronomy \& Astrophysics, Pennsylvania State University, University Park, PA 16802, USA}
\affil{\altaffilmark{5}Space Telescope Science Institute, 3700 San Martin Dr., Baltimore, MD 21218 USA}
\affil{\altaffilmark{6}Infrared Processing and Analysis Center, California Institute of Technology, Pasadena, CA 91125, USA}
\affil{\altaffilmark{7}NASA Ames Research Center, Moffett Field, CA 94035, USA}
\affil{\altaffilmark{8}Steward Observatory, University of Arizona, 933 North Cherry Avenue, Tucson, AZ 85721, USA}
\affil{\altaffilmark{9}Center for Astrophysics and Space Astronomy, University of Colorado, 389 UCB, Boulder, CO 80309, USA}
\affil{\altaffilmark{10}Minnesota Institute for Astrophysics, School of Physics and Astronomy, 116 Church Street, S. E., University of Minnesota, Minneapolis, MN 55455, USA}
\affil{\altaffilmark{11}Department of Physics, Florida State University, 77 Chieftain Way, Tallahassee, FL, 32306, USA}
\affil{\altaffilmark{12}MIT-Kavli Institute for Astrophysics and Space Research, 70 Vassar Street., Cambridge, MA 02139}
\affil{\altaffilmark{13}NSF Graduate Fellow}

\begin{abstract}
SPIRITS---SPitzer InfraRed Intensive Transients Survey---is an ongoing survey of nearby galaxies searching for infrared (IR) transients with \textit{Spitzer}/IRAC. We present the discovery and follow-up observations of one of our most luminous ($M_{[4.5]} = -17.1\pm0.4$~mag, Vega) and red ($[3.6] - [4.5] = 3.0 \pm 0.2$~mag) transients, SPIRITS\,15c. The transient was detected in a dusty spiral arm of IC~2163 ($D\approx35.5$~Mpc). Pre-discovery ground-based imaging revealed an associated, shorter-duration transient in the optical and near-IR (NIR). NIR spectroscopy showed a broad ($\approx 8400$~km~s$^{-1}$), double-peaked emission line of He~\textsc{i} at $1.083~\mu$m, indicating an explosive origin. The NIR spectrum of SPIRITS\,15c is similar to that of the Type IIb SN~2011dh at a phase of $\approx 200$~days. Assuming $A_V = 2.2$~mag of extinction in SPIRITS\,15c provides a good match between their optical light curves. The IR light curves and the extreme $[3.6]-[4.5]$ color cannot be explained using only a standard extinction law. Another luminous ($M_{4.5} = -16.1\pm0.4$~mag) event, SPIRITS\,14buu, was serendipitously discovered in the same galaxy. The source displays an optical plateau lasting $\gtrsim 80$~days, and we suggest a scenario similar to the low-luminosity Type~IIP SN~2005cs obscured by $A_V \approx 1.5$~mag. Other classes of IR-luminous transients can likely be ruled out in both cases. If both events are indeed SNe, this may suggest $\gtrsim 18\%$ of nearby core-collapse SNe are missed by currently operating optical surveys.
\end{abstract}


\section{Introduction} \label{sec:intro}
In the last few decades, the study of astrophysical transients has been revolutionized by the introduction of all-sky, high cadence surveys dedicated to their discovery. The largest advances have been made in the optical where the majority of time-domain surveys operate, but the dynamic infrared (IR) sky is only now beginning to be explored. IR follow-up of optically discovered transients has revealed new classes of events that can be dominated by IR emission, especially at late times. At least two known classes of transients with peak luminosities between those typical of novae and SNe can develop IR-dominated spectral energy distributions (SEDs) as they evolve:
\begin{inparaenum}[1)]
	\item stellar mergers, or luminous red novae, e.g., V1309 Sco \citep{tylenda11}, V838 Mon \citep{bond03, sparks08}, the 2011 transient in NGC~4490 (hereafter NGC~4490-OT, \citealp{smith16}), and M101~OT2015-1 (M101-OT, \citealp{blagorodnova16}),
	\item SN~2008S-like events, or intermediate luminosity red transients \citep{prieto08, thompson09, kochanek11}, also including NGC~300~OT2008-1 (hereafter NGC~300-OT, \citealp{bond09, humphreys11}), and PTF\,10fqs \citep{kasliwal11}.
\end{inparaenum}
Furthermore, otherwise luminous optical sources such as SNe may suffer extinction from obscuring dust, lending themselves to discovery and follow-up at IR wavebands where the effect of dust extinction is significantly reduced. 

Previous searches for obscured SNe have thus been motivated by the notion that if a significant fraction of SNe are heavily obscured, measurements of the SN rate from optical searches will only be lower limits \citep[e.g.][]{grossan99, maiolino02, cresci07}. Searches at near-IR (NIR) wavelengths have focussed on the dense, highly star-forming, nuclear regions of luminous infrared galaxies (LIRGS) and ultra-luminous infrared galaxies (ULIRGs) \citep[e.g.][]{vanburen94,grossan99,maiolino02,mannucci03,mattila05a,mattila05b,fox15}. These searches have had variable success, largely limited by insufficient angular resolution to probe the densest regions of starburst galaxies. 

High angular resolution studies using space-based telescopes or adaptive optics have found several candidates and 4 confirmed obscured SNe \citep{cresci07,mattila07,kankare08,kankare12}. Radio observations have also allowed the discovery of a few obscured SNe (SNe~II) in dense  star-forming regions, e.g., an SN in the starburst galaxy Mrk~297 \citep{yin91}, and SN~2008iz in M82 ($A_V > 10$~mag; \citealp{brunthaler09, brunthaler10, mattila13}). The SN rate estimates from such searches are still a factor of 3--10 lower than is expected from the high star formation rates inferred from the far-IR luminosities of the surveyed galaxies \citep[e.g.][]{cresci07}.

It has also been suggested that even in ``normal'' star-forming galaxies in the nearby universe, where extinction is much less extreme, the measured rates of core-collapse SNe (CCSNe) are still low compared to those expected from star formation rates \citep{horiuchi11}. This indicates that optical surveys may be missing populations of nearby SNe that are either intrinsically faint or hidden by dust. Moderate levels of visual extinction ($A_V \sim \mathrm{few}$~mag) in the less extreme star-forming environments of such galaxies are sufficient to dim some SNe beyond the detection limits of current optical surveys, further motivating IR transient searches of such hosts. 

Since December 2013, we have been conducting a systematic search for transients in the infrared (IR) with the SPitzer InfraRed Intensive Transients Survey (SPIRITS; PID11063; PI M. Kasliwal). This is an ongoing, 3-year targeted survey of 194 galaxies within 20~Mpc using the InfraRed Array Camera (IRAC; \citealp{fazio04}) aboard the \textit{Spitzer Space Telescope} (\textit{Spitzer}; \citealp{werner04,gehrz07}) at $3.6$ and $4.5~\mu$m ([3.6] and [4.5], respectively). Every galaxy in our sample has archival \textit{Spitzer}/IRAC imaging, such that our observing cadence covers time baselines from one week to several years. In our first year, SPIRITS discovered over 1958 infrared variable stars, and 43 transients. Four of these transients were in the luminosity range consistant with classical novae, and 21 were known SNe (for details see \citealp{johansson14}, \citealp{fox16}, and \citealp{tinyanont16}). 14 were in the IR luminosity gap between novae and supernovae and had no optical counterparts, possibly constituting a newly discovered class of IR-dominated transients (Kasliwal et al. 2016, submitted to ApJ).

Here we report the discovery of two transients discovered in dusty spiral arms of the galaxy IC~2163, SPIRITS\,15c and SPIRITS\,14buu. The IR luminosity of SPIRITS\,15c was brighter than $-17$~mag, one of the most luminous transients discovered by SPIRITS to date and more luminous than the new classes of IR-dominated transients discussed above. Additionally, the spectrum of SPIRITS\,15c is dominated by a broad emission line of He~\textsc{i}, suggesting this is an explosive event such as an SN, but with significant dust extinction that obscured the transient in the optical. In \S~\ref{sec:discovery}, we describe the discovery and optical/IR follow up observations of SPIRITS\,15c, and the subsequent post-outburst, serendipitous discovery of SPIRITS\,14buu. In \S~\ref{sec:analysis}, we describe the analysis of our photometric and spectroscopic data. In \S~\ref{sec:discussion}, we explore the possibility that SPIRITS\,15c is an obscured SN based on the similarity of its NIR spectrum to that of the well studied Type IIb SN~2011dh. Using a similar analysis we consider that SPIRITS\,14buu is yet another obscured SN, likely of Type IIP. We also consider non-supernova IR transient scenarios in \S~\ref{sec:non-SN}, including stellar mergers, SN~2008S-like events, and the proposed helium nova V445 Pup. Finally, in \S~\ref{sec:conclusions}, we summarize the observational characteristics of SPIRITS\,15c and SPIRITS\,14buu and present our conclusions.

\section{SPIRITS discovery and follow-up observations} \label{sec:discovery}
We present the discovery of SPIRITS\,15c, and describe observations of this event from our concomitant ground-based survey of SPIRITS galaxies. Additionally, in \S~\ref{sec:IC2163OT}, we discuss the serendipitous discovery of another, earlier transient in the same galaxy, SPIRITS\,14buu. 

\subsection{\textit{Spitzer}/IRAC Discovery in IC\,2163}\label{sec:Spitzer}
\begin{figure*}
\centering
\begin{minipage}{180mm}
\includegraphics[width=\linewidth]{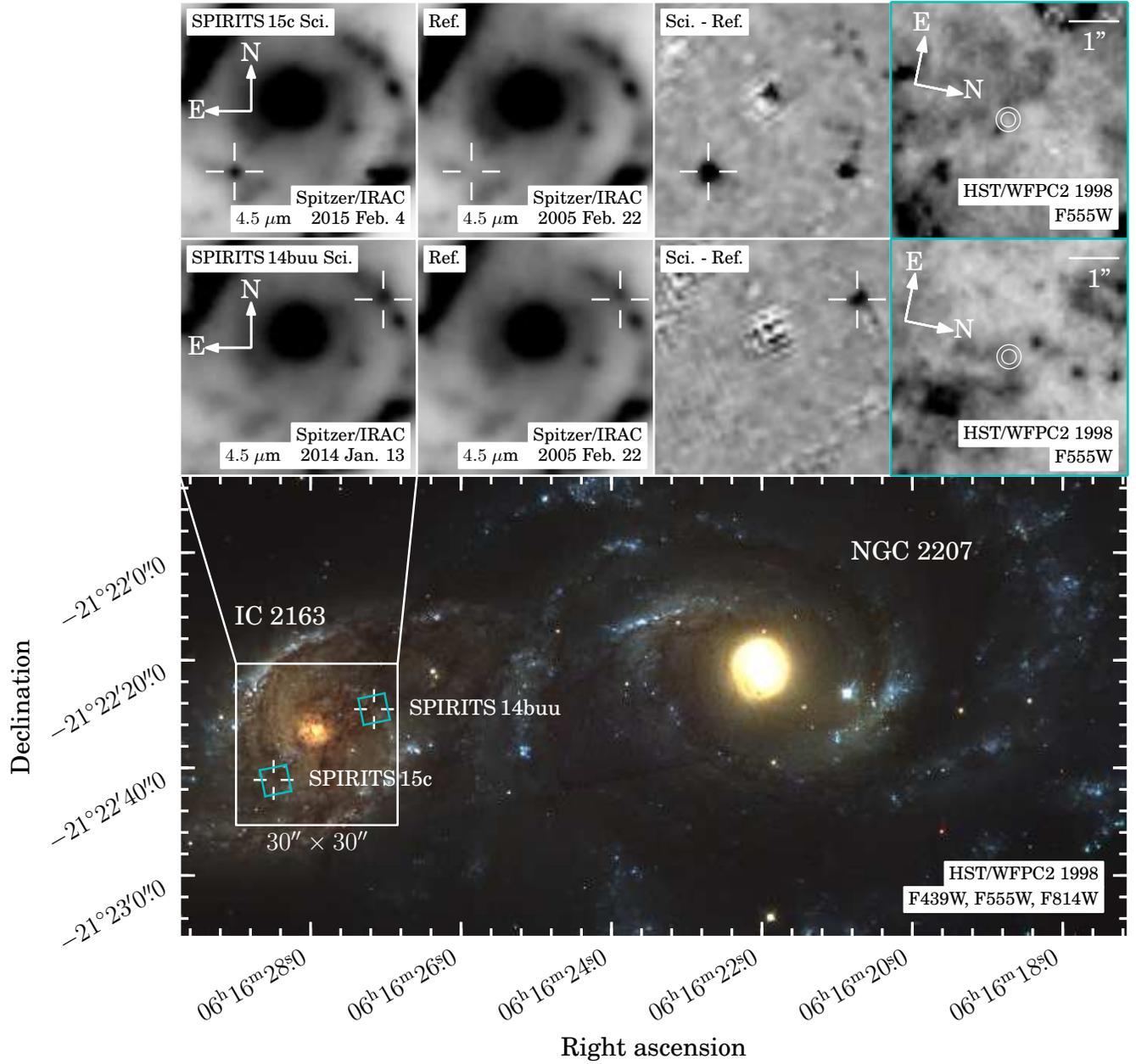}
\caption
{ \label{fig:Fig1}
The bottom panel shows color-composite, archival \textit{HST}/WFPC2 imaging of the interacting galaxy pair of IC~2163 and NGC~2207 from 1998 in three filters (F439W in blue, F555W in green, and F814W in red). The positions of SPIRITS\,15c and SPIRITS\,14buu in dusty spiral arms of IC~2163 are indicated with the crosshairs. The right panels of the top and middle rows show zoom-ins of the F555W image to the regions indicated by the cyan squares for SPIRITS\,15c and SPIRITS\,14buu, respectively. The white circles show the 3- and 5-$\sigma$ uncertainties on the position of transients. We are unable to identify an individual star as a candidate progenitor in either case. The left three panels in the top and middle rows show the $30''\times30''$ region indicated by the white zoom-in box. The first panels from the left are the \textit{Spitzer}/IRAC discovery science frames at [4.5] from 2015 February 4.4 (top row; SPIRITS\,15c) and 2014 January 13.9 (middle row; SPIRITS\,14buu), the second are the reference images from 2005 February 22.7 (PID3544; PI D.~M. Elmegreen), and the third are the $\mathrm{science} - \mathrm{reference}$ subtraction images, clearly showing the new, transient sources. We note in the SPIRITS\,15c discovery and subtraction images that the apparent variability of the galaxy nucleus and the apparent new source directly to the west of SPIRITS\,15c are likely spurious.}
\end{minipage}
\end{figure*}

As part of the regular SPIRITS observing program, the interacting pair of late-type galaxies IC~2163 and NGC~2207 was observed at 10 epochs between 2014 January 13.9 and 2016 January 12.1 using \textit{Spitzer}/IRAC at [3.6] and [4.5]. Image subtraction was performed using archival \textit{Spitzer}/IRAC images from 2005 February 22.7 as references. For details on our image subtraction pipeline see Kasliwal et al. (2016). On 2015 February 19, a transient source, designated SPIRITS\,15c, was identified by the SPIRITS team in an image at [4.5] taken on 2015 February 4.4.\footnote{The lag between the observations and discovery of SPIRITS\,15c was due to the time for \textit{Spitzer}/IRAC observations to be made available on the \textit{Spitzer} Heritage Archive. Since 2015 May, the SPIRITS team has made extensive use of the \textit{Spitzer} Early Release Data Program, allowing us to discovery transients within a few days of observation.} The [4.5] discovery image are shown in Figure~\ref{fig:Fig1}. 

SPIRITS\,15c was discovered at a right ascension and declination of $06^{\mathrm{h}}16^{\mathrm{m}}28\fs49, -21\degree22\arcmin42\farcs2$ (J2000), coincident with a spiral arm of the galaxy IC~2163. We assume a distance modulus to IC~2163 of $\mu = 32.75 \pm 0.4$~mag ($\approx 35.5$~Mpc, \citealp{theureau07} from NED\footnote{The NASA/IPAC Extragalactic Database (NED) is operated by the Jet propulsion Laboratory, California Institute of Technology, under contract with the National Aeronautics and Space Administration.}; $z=0.00922$, $v=2765$~km~s$^{-1}$, \citealp{elmegreen00}). We note that IC~2163 was selected as part of the SPIRITS sample based on an incorrect distance estimate that placed it within the 20~Mpc cutoff for SPIRITS galaxies. The foreground Galactic extinction toward IC~2163 is $A_V = 0.238$~mag from NED (\citealp{fitzpatrick99, schlafly11}). There is most likely additional extinction from the foreground spiral arm of NGC~2207.

\begin{figure*}
\centering
\begin{minipage}{180mm}
\includegraphics[width=\linewidth]{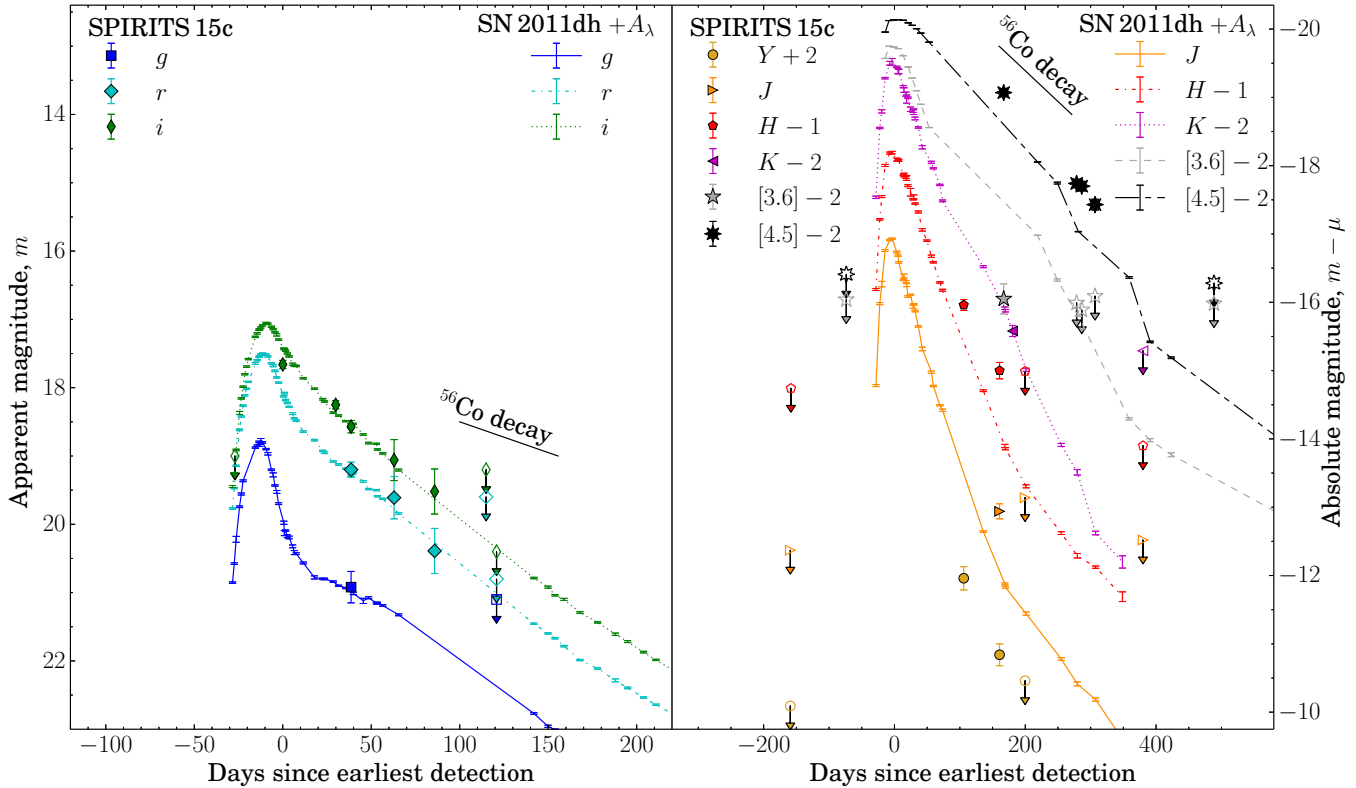}
\caption
{ \label{fig:Fig2}
The optical ($gri$; AB magnitudes; left panel), and IR ($YJHK$, [3.6], and [4.5]; Vega magnitudes; right panel) light curves of SPIRITS\,15c (points) and the Type IIb SN~2011dh (lines). Unfilled points with downward arrows indicate upper limits. Error bars are shown but are sometimes smaller than the points. Time on the $x$-axis is given as the number of days since the earliest detection of SPIRITS\,15c on 2014 August 21.4 ($\mathrm{MJD} = 56890.4$). The light curves of SN~2011dh are shifted in apparent magnitude to the distance of SPIRITS\,15c, with applied total reddening characterized by $E(B-V)_{\mathrm{MW}} + E(B-V)_{host} = 0.72$~mag, or $A_V \approx 2.2$ magnitudes of extinction assuming a standard $R_V=3.1$ extinction law \citep{fitzpatrick99, chapman09, schlafly11}. The phase of the SN~2011dh light curves is set so that the earliest detection of SN~2011dh coincides with the most constraining non-detection preceding the outburst of SPIRITS\,15c. For comparison, we also show the expected decay rate for a light curve powered by the radioactive decay of $^{56}$Co (see \citealp{gehrz88,gehrz90}).} 
\end{minipage} 
\end{figure*}

Photometry was performed at the position of SPIRITS\,15c on the reference subtracted images using a 4 mosaicked pixel aperture and a background annulus from 4--12 pixels. The extracted flux was multiplied by aperture corrections of 1.215 and 1.233 for [3.6] and [4.5], respectively, as outlined in the IRAC instrument handbook. The photometric measurements, given in Table~\ref{table:phot}, indicate a high luminosity of $M_{[4.5]} = -17.1 \pm 0.4$~mag at the distance of IC~2163 and an extremely red color of $[3.6] - [4.5] = 3.0 \pm 0.2$~mag at the epoch of discovery. SPIRITS\,15c was not detected in earlier images taken on 2014 Jun 8.7 to a limiting magnitude of $m\gtrsim18.3$~mag at [4.5] ($-14.5$~mag absolute), providing an age constraint for SPIRITS\,15c as an active mid-IR (MIR) source of $240.7$~days. Additional detections were made at [4.5] in SPIRITS observations of IC~2163 on 2015 May 26.9, June 3.8, and June 24.1, fading by $\approx 2$~magnitudes in 140~days. On 2015 December 23.0, SPIRITS\,15c was undetected at [4.5] and remained so in all subsequent SPIRITS observations through the most recent one on 2016 January 12.1. The source was undetected at [3.6] in all post-discovery epochs of \textit{Spitzer}/IRAC observation. The 5-$\sigma$ depth in the reference images at the position of SPIRITS\,15c is 14.9~mag ($\approx -17.9$~mag absolute) in both bands, and thus, we are unable to place meaningful constraints on the IR properties of the progenitor system. The full sequence of photometric measurements from \textit{Spitzer}/IRAC for SPIRITS\,15c are summarized in Table~\ref{table:phot} and shown in Figure~\ref{fig:Fig2} along with the NIR and optical measurements described below in \S~\ref{sec:ground_imaging}.  

\subsubsection{Host Galaxy Properties}

\citet{elmegreen16} report global values of the $24~\mu\mathrm{m}$ flux density from a \textit{Spitzer}/MIPS 2013 archival image of $S_{\nu}(24~\mu\mathrm{m}) = 5.9\times10^2$~mJy for IC~2136, and $2.06\times10^{4}$~mJy for the galaxy pair as a whole. They also give global H$\alpha$~fluxes, derived from narrowband H$\alpha$ and broadband $R$ images from \citet{elmegreen01}, of $S(\mathrm{H}\alpha) = 5.9\times10^{-13}$ and $3.49\times10^{-12}$~erg~s$^{-1}$~cm$^{-2}$ for IC~2163 and the pair as a whole, respectively. Combining these and following \citet{kennicutt09} give global star formation rates of $1.9~M_{\odot}$~yr$^{-1}$ for IC~2163 and $6.8~M_{\odot}$~yr$^{-1}$ for both galaxies combined. The rate of SNe associated with this galaxy pair is also high, with 5 known SNe hosted in IC~2163 and NGC~2207 since 1975. These include four CCSNe (SN~1999ec, \citealp{modjaz99,jha99}; SN~2003H, \citealp{filippenko03}; SN~2013ai, \citealp{conseil13}; SN~2010jp, \citealp{smith12}) and one thermonuclear SN (SN~1975A, \citealp{kirschner76}). The most recent SN associated with the galaxy pair, SN~2010jp, a peculiar SN~IIn with evidence for strong interaction with a dense circumstellar medium (CSM) and a bipolar, jet-driven explosion \citep{smith12}. 

In a $3.6$~arcsec aperture centered on the location of SPIRITS\,15c, $S_{\nu}(24~\mu\mathrm{m}) = 8.38$~mJy, measured in the HiRes devconvolution image from \citet{velusamy08} ($1.9$~arcsec resolution). In the same aperture, $S(\mathrm{H}\alpha) = 1.25\times10^{-14}$~erg~s$^{-1}$~cm$^{-2}$. Addtionally, the $^{12}$CO(1-0) (115.27~GHz, 2.6~mm) line flux in this region, measured with ALMA, implies a mean line of sight H$_2$ surface density of $\Sigma_{\mathrm{H}_2} = 13.7~M_{\odot}$~pc$^{-2}$ (\citealp{elmegreen16}, private communication). Estimates of the star formation rate using H$\alpha$ as tracer in such a small aperture will include stochastic effects related to including too few O stars, but these measurements suggest on-going star formation in the vinicity of SPIRITS\,15c and that the transient may be associated with a young stellar population. The heliocentric radial velocity in the ALMA CO velocity image $v_{\mathrm{CO},helio}=2827$~km~s$^{-1}$ at the nearest non-blanked pixel to SPIRITS\,15c is $v_{\mathrm{CO},\rm{helio}}=2827$~km~s$^{-1}$ ($2''$ separation; \citealp{elmegreen16}, private communication). We adopt this value as the velocity of SPIRITS\,15c with respect to the observer frame throughout this paper. 

\subsection{Ground-based Imaging}\label{sec:ground_imaging}
To supplement our \textit{Spitzer}/IRAC observations, the SPIRITS team is undertaking a concomitant, ground-based survey of SPIRITS galaxies in the optical and NIR using a number of telescopes and instruments. 

IC~2163 was imaged at several epochs in $gri$ with the CCD camera on the Swope Telescope at Las Campanas Observatory (LCO). NIR $YJH$ imaging of SPIRITS\,15c was also obtained at several epochs with the RetroCam infrared camera on the Du Pont Telescope, and $J$, $H$, and $K_s$-band images were obtained with the FourStar IR camera on the Magellan Baade Telescope at LCO. Photometry was extracted using simple aperture photometry, with the aperture size defined by the seeing in each image, and calibrated against SDSS field stars for the optical images and 2MASS stars for the NIR images. For $Y$-band images, we adopt the conversion from 2MASS used for the Wide Field Camera on the United Kingdom Infrared Telescope from \citet{hodgkin09}. Image subtraction was performed for the Swope/CCD and Du Pont/Retrocam images using template images taken on 2015 Mar 13.2 (after SPIRITS\,15c had faded) and 2014 Jan 10.2, respectively, to obtain more accurate photometry and deeper limits from non-detection images.

We also obtained optical imaging with the Las Cumbres Observatory Global Telescope (LCOGT) Network 1-m telescopes in the $r$- and $i$-bands. Photometry was performed by simultaneously fitting the point spread function (PSF) of the transient, measured using SDSS field stars, and background, modelled using low order polynomials. The photometric measurements were also calibrated against SDSS field stars.

Additional imaging of IC~2163 was obtained on 2014 Feb 6.2 and 2015 Jan 20.0 with the Inamori Magellan Areal Camera and Spectrograph (IMACS) on the Magellan Baade Telescope using a wide-band, red filter (WB6226-7171 calibrated to SDSS $r^{\prime}$), on 2015 March 16 using the Mont4k on the Kuiper 61'' Telescope in $R$-band, on 2013 December 24.3 with the optical camera on the 60-in telescope at Palomar observatory (P60) in $g$, $r$, and $i$, and on 2013 December 16.1 in $K_s$-band with the Nordic Optical Telescope near-infrared Camera and spectrograph (NOTCam) at the Observatorio del Roque de los Muchachos.

The earliest detection of SPIRITS\,15c on 2014 August 21.4 (MJD = 56890.4) was also the observed optical peak at $i=17.66 \pm 0.06$~mag. Correcting for Galactic extinction gives a peak absolute magnitude $M_{i} \leq -15.1 \pm 0.4$~mag, where any additional foreground or host extinction would imply a higher intrinsic luminosity. Throughout the rest of this paper, we refer to days $t$ since the earliest detection of SPIRITS\,15c on this date. Our most stringent constraint on the age of SPIRITS\,15c is the 28.4 day window between the earliest detection and the previous optical non-detection on 2014 July 24.0. The observed optical colors at $t=38.6$~days were quite red with $g-r = 1.71 \pm 0.15$~mag and $g-i = 2.24 \pm 0.15$~mag. By $t=120.8$~days, SPIRITS\,15c had faded beyond detection with Swope/CCD, but was still detected in the Baade/IMACS image at $21.4 \pm 0.1$~mag.

At $t = 105.9$~days, SPIRITS\,15c was detected in the NIR with $H = 17.84 \pm 0.06$~mag ($-14.91 \pm 0.4$~mag absolute), and a red NIR color with $Y-H = 1.04 \pm 0.09$~mag. The transient then faded in the NIR by $1$~mag at nearly constant color over the next $54.8$~days. All of our photometric measurements for SPIRITS\,15c are summarized in Table~\ref{table:phot} and the light curves are shown in Figure~\ref{fig:Fig2}.

\subsection{Another transient discovery: SPIRITS\,14buu}\label{sec:IC2163OT}
During the analysis of our imaging data, a second transient was serendipitously noticed in IC~2163 at a right ascension and declination of $06^{\mathrm{h}}16^{\mathrm{m}}27\fs2, -21\degree22\arcmin29\farcs2$ (J2000), located in a spiral arm directly on the other side of the galaxy from SPIRITS\,15c. This transient was active in January 2014 SPIRITS \textit{Spitzer}/IRAC data before the implementation of our automated image subtraction and candidate identification pipeline. It was retroactively assigned the name SPIRITS\,14buu. We show the [4.5] discovery image in Figure~\ref{fig:Fig1}.

\begin{figure*}
\centering
\centering
\begin{minipage}{180mm}
\includegraphics[width=\linewidth]{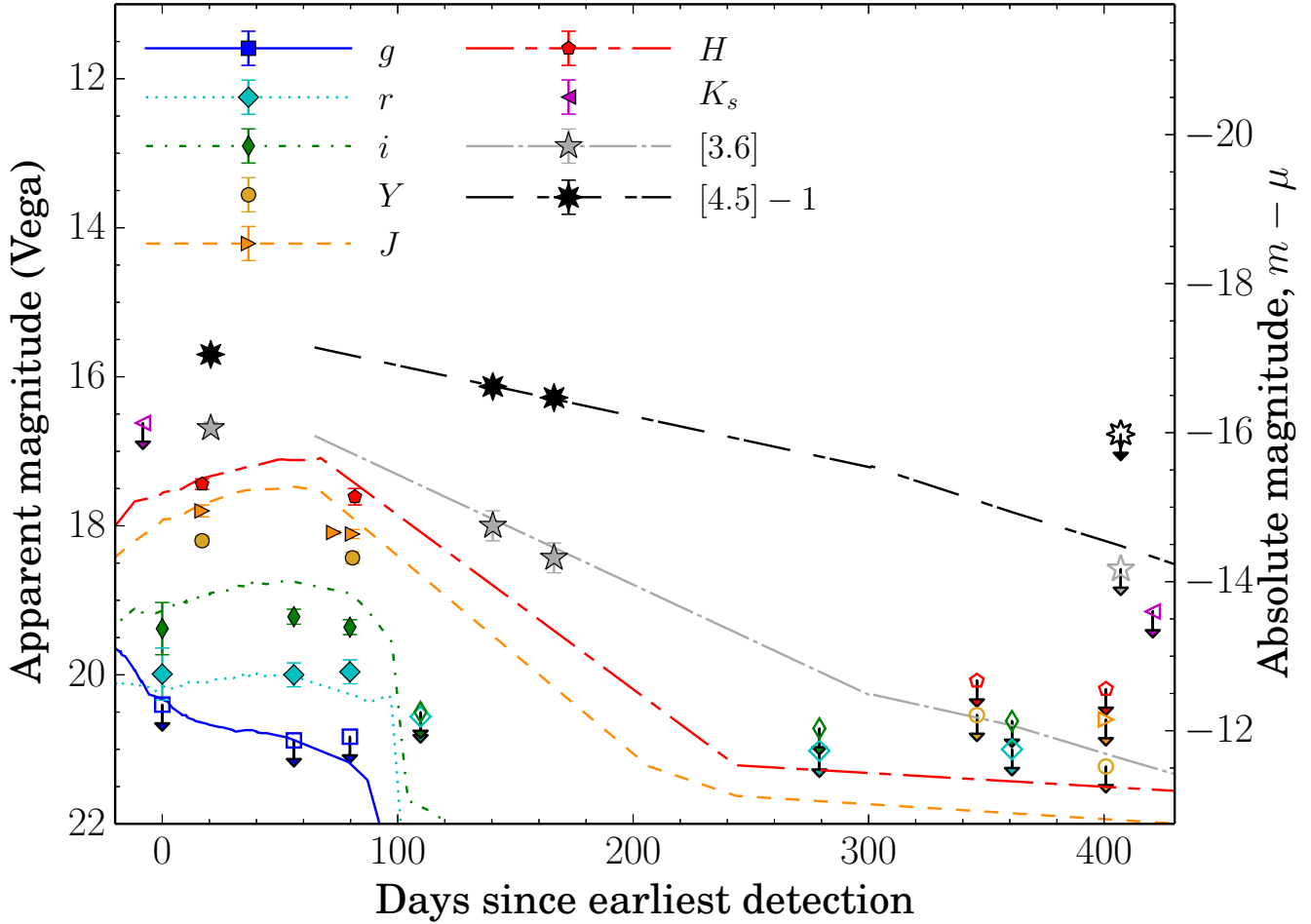}
\caption
{ \label{fig:Fig3}
The optical ($gri$; AB magnitudes), and IR ($YJH$, [3.6], and [4.5]; Vega magnitudes) light curves of SPIRITS\,14buu (points). Also shown are the $griJH$ light curves of the Type IIP SN~2005cs and the [3.6] and [4.5] light curves of SN~2004et (lines). Unfilled points with downward arrows indicate upper limits. Error bars are shown, but are usually smaller than the points. Time on the $x$-axis is given as the number of days since the earliest detection of SPIRITS\,14buu on $\mathrm{MJD} = 56650.3$. The light curves of SN~2005cs are shifted in absolute magnitude by 0.1~mag, with applied total reddening characterized by $E(B-V)= 0.49$, or $A_V \approx 1.5$~mag assuming a standard $R_V$=3.1 extinction law \citep{fitzpatrick99, chapman09, schlafly11}. The light curves of SN~2004et are shifted fainter by $2.4$~mag. The phase of the SN~2005cs light curves is set so that fall-off of the plateau lines up with that of SPIRITS\,14buu. The phase of the SN~2004et light curves is time in days from the assumed explosion epoch ($\mathrm{MJD} = 53270.0$; \citealp{li05}).}
\end{minipage}
\end{figure*}

The earliest detection of this event is in the P60 $r$- and $i$-band images taken on 2013 December 24.3 (MJD = 56650.3). It was not detected in the $K_s$ NOTCam image on 2013 December 16.1 to $K_s > 16.6$~mag ($> -16.2$~mag absolute), but this limit is not deep enough to constrain the age of SPIRITS\,14buu. The optical and IR light curves of SPIRITS\,14buu are given in Table~\ref{table:phot_14buu} and shown in Figure~\ref{fig:Fig3}. In the $r$- and $i$-bands, the source displays a plateau lasting at least $80$~days at $i = 19.2 \pm 0.1$~mag ($M_i = -13.7$~mag absolute) and with $r-i = 0.7 \pm 0.2$~mag. The source is undetected in $g$-band indicating $g-i \gtrsim 1.7$~mag. Following the plateau, the optical light curves fall off steeply, dropping in $i$-band by $\gtrsim 1.1$~mag in $\lesssim 24$~days. A luminous NIR counterpart was detected during the optical plateau with $H = 17.44 \pm 0.07$ ($M_H = -15.3$ absolute). The source shows red NIR colors with $Y-H = 0.76 \pm 0.09$ and $J-H = 0.4 \pm 0.1$~mag, and a slow decline in the NIR of $\approx 0.2$~mag over 65~days.

SPIRITS\,14buu was also detected as a luminous MIR source with $Spitzer/IRAC$. The observed MIR peak occurred at 17~days after the earliest detection of this event at $16.7 \pm 0.1$~mag ($-16.1$ absolute) in both bands. At [4.5], the light curve declines slowly by $0.6 \pm 0.1$~mag over $145.8$~days, extending well past the time of the plateau fall-off in the optical. At [3.6], the source fades more quickly developing a red $[3.6] - [4.5]$ color of $1.2 \pm 0.2$~mag by 166.4 days after the earliest detection. The source faded beyond detection in both bands by $407.1$~days.

We further describe the characteristics of this transient and discuss possible physical interpretations below in \S~\ref{sec:IC2163OT_SN}. 

\subsection{Archival \textit{HST} Imaging\footnote{Based on observations made with the NASA/ESA Hubble Space Telescope, obtained from the data archive at the Space Telescope Science Institute. STScI is operated by the Association of Universities for Research in Astronomy, Inc. under NASA contract NAS 5-26555.}}\label{sec:hst}
Images of the interacting galaxies IC~2163 and NGC~2207 were obtained with the {\it Hubble Space Telescope\/} ({\it HST}) on 1998 November~11 in program GO-6483 (PI: D.~Elmegreen). These observations used the Wide Field Planetary Camera~2 (WFPC2), and fortuitously covered the site of SPIRITS\,15c and SPIRITS\,14buu, some 17 years before their outbursts. Images of this galaxy pair were also included in the Hubble Heritage collection. In the bottom panel of Figure~\ref{fig:Fig1}, we show an RGB color-composite, mosaicked image of the field containing the transients made from imaging data in the F814W (red), F555W (green), and F439W (blue) filters. We note the locations of SPIRITS\,15c and SPIRITS\,14buu in dusty spiral arms of IC~2163.

In the \textit{HST} images of IC~2163, the location of SPIRITS\,15c unfortunately lies close to the edge of the PC chip, so we did not further analyze those frames. In the images of NGC~2207 the SPIRITS\,15c site is well placed in the WF2 field. We chose two sets of dithered frames taken with the F555W filter ($\sim V$; exposures of $3\times160$~s plus one of 180~s), and with F814W ($\sim I$; $4\times180$~s). We registered and combined these two sets of images, using standard tasks in IRAF\footnote{IRAF is distributed by the National Optical Astronomy Observatory, which is operated by the Association of Universities for Research in Astronomy (AURA) under a cooperative agreement with the National Science Foundation.}.

To find the precise location of SPIRITS\,15c in the WFPC2 frames, we employed the Baade/IMACS WB6226-7171 image from 2015 January 20.0. Using centroid measurements for 15 stars detected both in the Baade/IMACS frame and in the two combined WFPC2 images, we determined the geometric transformation from Magellan to WFPC2 using the STSDAS\footnote{STSDAS (Space Telescope Science Data Analysis System) is a product of STScI, which is operated by AURA for NASA.} {\tt geomap} task. By applying the {\tt geotran} task to the Magellan frame, and blinking this transformed image against the WFPC2 frames, we verified the quality of the registration. Then the {\tt geoxytran} task yielded the $x,y$ pixel location of SPIRITS\,15c in the {\it HST\/} images. The standard deviations of the geometric fits for the reference stars were 0.50 and 0.48 pixels in each coordinate in F555W and F814W, respectively, corresponding to $0.05''$.

The precise location of SPIRITS\,15c is shown in the upper right panel of Figure~\ref{fig:Fig1} with 3- and 5-$\sigma$ error circles overlaid on the F555W image. At the pre-eruption site in 1998 we detect no progenitor star in either F555W or F814W. The location is within a dark dust lane, with very few stars detected in its vicinity. The 5-$\sigma$ limiting magnitudes in the {\it HST\/} frames are $V>25.1$ and $I>24.0$~mag. These correspond to absolute magnitudes of $M_V > -7.9$ and $M_I > -8.9$~mag at the assumed distance of IC~2163, correcting only for Galactic extinction. 

We preformed the same analysis on the Baade/IMACS image from 2014 Feb 6.2 with a clear detection of SPIRITS\,14buu to analyze the location of this transient. Again, we found the standard deviations of the geometric fits for the reference stars were $\approx 0.50$ pixels ($\approx 0.05''$) in each coordinate in both the F555W and F814W frames. The precise location of SPIRITS\,14buu is shown in the right panel of Figure~\ref{fig:Fig1}. The transient is coincident with a poorly resolved stellar association or cluster in the center of a dusty spiral arm, but it is not possible to identify an individual star as a candidate progenitor given the distance to IC~2163. 

\subsection{Near-IR Spectroscopy}
We obtained an epoch of NIR spectroscopy of SPIRITS\,15c with the Folded-port InfraRed Echellette spectrograph (FIRE; \citealp{simcoe08,simcoe13}) on the Magellan Baade Telescope at LCO at $t=205$~days on 2015 March 14 and a later epoch with the Multi-Object Spectrometer for Infra-Red Exploration (MOSFIRE; \citealp{mosfire10,mosfire12}) on the Keck I Telescope at W.~M. Keck Observatory at $t=222$~days on 2015 March 31. SPIRITS\,14buu was also included in the MOSFIRE slit mask, but the source was not detected above the host galaxy light. This was 60 days after the du Pont/Retrocam non-detection of SPIRITS\,14buu at $J>20.6$~mag. 

The Baade/FIRE observations were obtained using the low dispersion, high throughput prism mode and a completed ABBA dither sequence. The data span 0.8--2.5~$\mu\mathrm{m}$ at a resolution ranging from 300--500. Immediately afterwards, we obtained a spectrum of the flux and telluric standard star HIP~32816. The data were reduced using the IDL pipeline \textsc{firehose}, which is specifically designed to reduce FIRE data \citep{simcoe13}. The pipeline performed steps of flat fielding, wavelength calibration, sky subtraction, spectral tracing, and extraction. The sky flux was modeled using off-source pixels as described by \citet{kelson03} and was subtracted from each frame. The spectral extraction was then performed using the optimal technique by \citet{horne86} to deliver the maximum signal-to-noise ratio while preserving spectrophotometric accuracy. Individual spectra were then combined with sigma clipping to reject spurious pixels. Corrections for telluric absorption were performed using the IDL tool \textit{xtellcor} developed by \citet{vacca03}. To construct a telluric correction spectrum free of stellar absorption features, a model spectrum of Vega was used to match and remove the hydrogen lines of the Paschen and Brackett series from the A0V telluric standard, and the resulting telluric correction spectrum was also used for flux calibration.​

\begin{figure*}
\centering
\begin{minipage}{180mm}
\includegraphics[width=\linewidth]{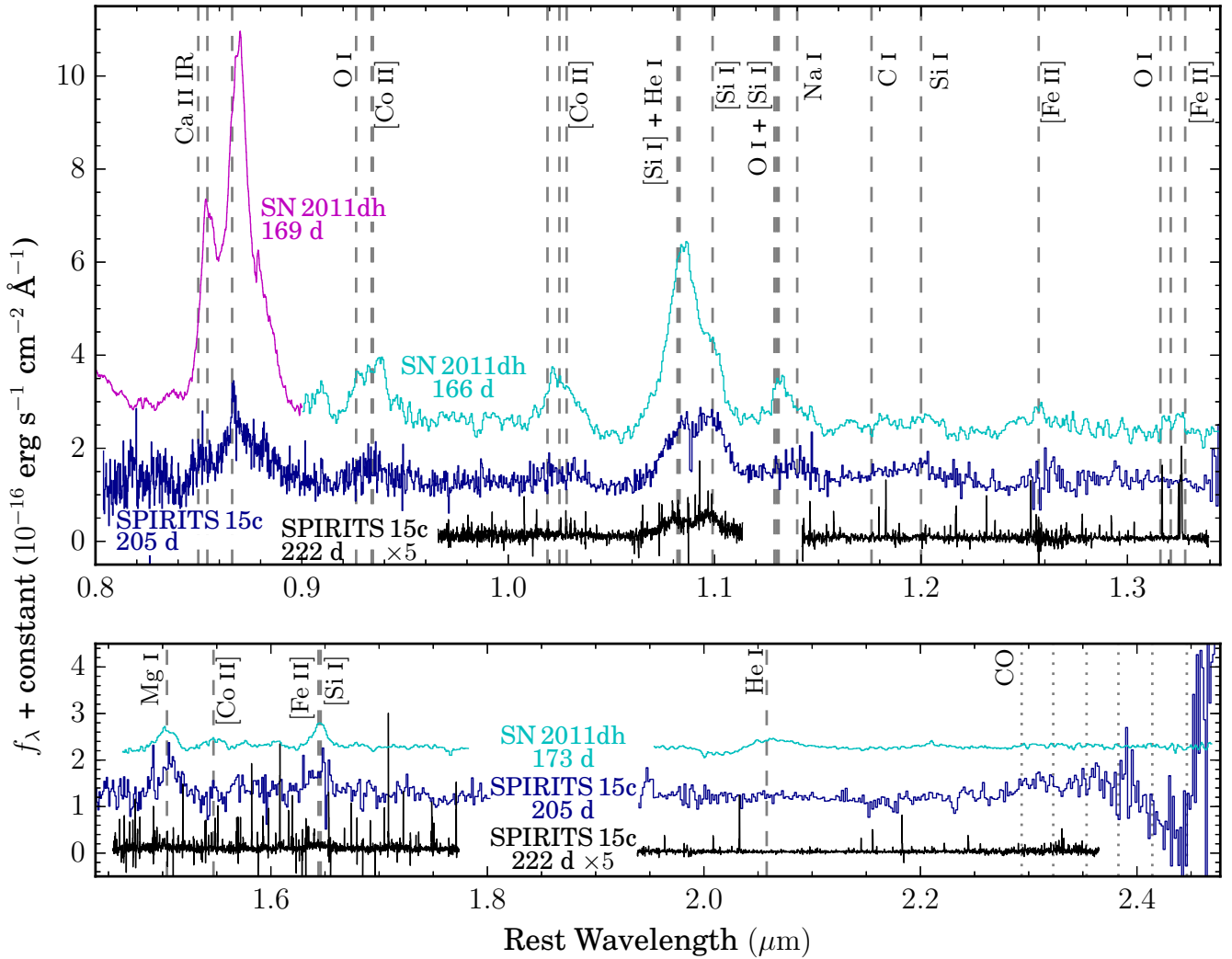}
\caption
{ \label{fig:Fig4}
The NIR spectra of SPIRITS\,15c taken with Baade/FIRE on $t=205$~days (blue) and Keck/MOSFIRE on $t=222$~days (black) shifted to the rest frame of the the location of SPIRITS\,15c measured in an ALMA CO velocity map of IC~2163 ($v_{\mathrm{CO},helio} = 2827$~km~s$^{-1}$, \citealp{elmegreen16}). The Keck/MOSFIRE spectrum is multiplied by a factor of 5 and the Baade/FIRE spectrum is shifted up by $10^{-16}$~erg s$^{-1}$ cm$^{-2}$ \AA$^{-1}$ for clarity. We also show late-time spectra of SN~2011dh for comparison (cyan and magenta), shifted up by $2\times10^{-16}$~erg s$^{-1}$ cm$^{-2}$ \AA$^{-1}$, where the phases marked on the plot are set as in Figure~\ref{fig:Fig2}. Dashed vertical lines show the locations of atomic transitions identified in the spectra of SN~2011dh by \citet{ergon14,ergon15} and \citet{jerkstrand15}. Dotted vertical lines mark the band heads of the $\Delta v = 2$ vibrational overtones of $^{12}$C$^{16}$O, which may contribute to excess flux observed between $2.3$ and $2.5~\mu$m.} 
\end{minipage}
\end{figure*}

The Keck/MOSFIRE observations were carried out using 180~s exposures in the $Y$- and $K$-bands and 120~s exposures in $J$ and $H$. The target was nodded along the slit between exposures to allow for accurate subtraction of the sky background. Total integration times in each band were 1431.5~s in $Y$, 715.8~s in $J$, 1192.9~s in $H$, and 1079.6~s in $K$. Spectral reductions, including flat-fielding, the wavelength solution, background subtraction, and frame stacking, were performed using the MOSFIRE Data Reduction Pipeline. 1D spectra were extracted from the 2D frames using standard tasks in \textsc{iraf}. The A0V telluric standard HIP~30090 was also observed immediately following the observations of SPIRITS\,15c and telluric corrections and flux calibrations were performed using \textit{xtellcor}. The 1D spectra are shown in Figure~\ref{fig:Fig4}, shifted to the rest frame of the $^{12}$CO ALMA velocity measurements at the position of SPIRITS\,15c discussed in \S~\ref{sec:Spitzer}.

\section{Analysis} \label{sec:analysis}
In this section, we provide our analysis of the observed spectral energy density (SED) evolution of SPIRITS\,15c and of its NIR spectrum. 

\subsection{The Optical to IR SED}
\begin{figure}
\centering
\includegraphics[width=\linewidth]{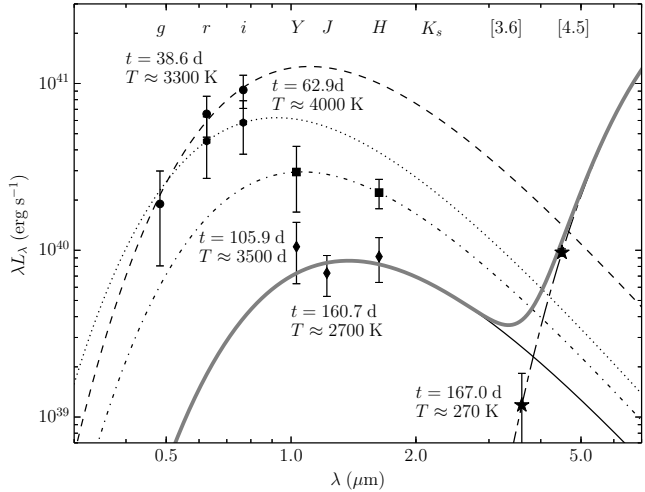}
\caption
{ \label{fig:Fig5}
The optical to IR SED of SPIRITS\,15c at multiple epochs during the evolution of the transient. Band-luminosities ($\lambda L_{\lambda}$), calculated assuming a distance to IC~2163 of $35.5$~Mpc and correcting for Galactic reddening characterized by $E(B-V) = 0.072$~mag, are shown at $t=38.6$ (circles), $62.9$ (hexagons), $105.9$ (squares), $160.7$ (diamonds), and $167.0$~days (stars). Best-fit, single-component blackbody curves are shown as black dashed, dotted, dash-dotted, solid, and long-short dashed lines, respectively, with their corresponding temperatures labeled. Since the final two epochs are nearly contemporaneous ($\Delta t = 6.3$~days), we show the sum of these two blackbody components as the solid gray curve.}
\end{figure}

We constructed a spectral energy distribution (SED) of SPIRITS\,15c at multiple epochs ($t=39, 63, 106$, $161$, and $167$~d) during the evolution of the transient using the available optical and IR photometry. The photometric points, converted to band-luminosities ($\lambda L_{\lambda}$) assuming a distance to the host galaxy of $35.5$~Mpc and correcting for Galactic reddening, are shown in Figure~\ref{fig:Fig5} along with blackbody fits to the data at each epoch. At $t=39$~days, the optical points are well approximated by a blackbody with $T\approx3300$~K and $R\approx1.4\times10^{15}$~cm. Fits to the NIR points seem to indicate the SED evolves in time to a higher effective temperature and smaller radius, with $T\approx3500$~K and $R\approx6.3\times10^{14}$~cm by $t=106$~days. At $t=161$--$167$~days, the SED shows at least two distinct components, with the NIR points approximated by a $T\approx2700$~K, $R\approx5.7\times10^{14}$~cm blackbody and a required cooler component to account for the [3.6] and [4.5] \textit{Spitzer}/IRAC measurements, likely associated with dust emission. A blackbody fit to the measured band-luminosities at [3.6] and [4.5] gives a color temperature of $T\approx270$~K. We note, however, that the sum of these two components significantly over-predicts the luminosity at [3.6]. 

Given that SPIRITS\,15c coincides with an apparently dusty region of IC~2163, the observed SED is likely affected by significant extinction from host galaxy dust. Thus, it is difficult to infer intrinsic properties of the transient from our photometric data. In Table~\ref{table:bb_fits}, we give the results of blackbody fits to the data assuming several different values for the total extinction. 
 
\subsection{The NIR spectrum}\label{sec:nir_spec}
\begin{figure}
\centering
\includegraphics[width=\linewidth]{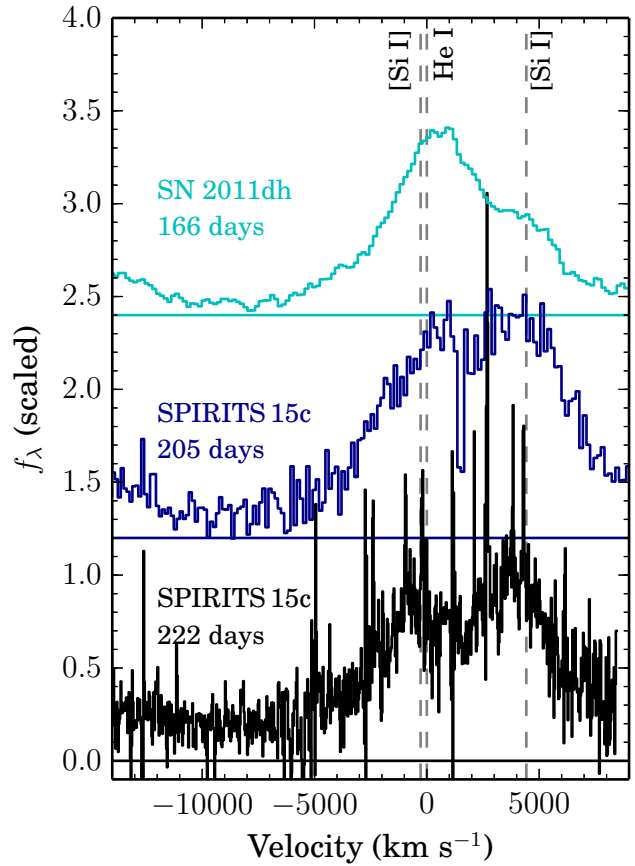}
\caption
{ \label{fig:Fig6}
The observed velocity profile of the He~\textsc{i} line at $1.0830~\mu$m in the spectra of SN~2011dh at a phase of $166$~days (top) and SPIRITS\,15c at phases of $205$ (middle) and $222$~days (bottom). The phase of the SN~2011dh spectrum is set as in Figure~\ref{fig:Fig2}. The profiles are scaled and shifted arbitrarily in flux for clarity.}  
\end{figure}

SPIRITS\,15c was observed spectroscopically in the NIR at $t=205$ and $222$~days during the IR decline phase of the transient. As seen in Figure~\ref{fig:Fig4}, a prominent feature in both spectra is the broad emission line near $1.083~\mu$m, likely due to He~\textsc{i}. The full-width at half-maximum (FWHM) velocity of this line at $t=166$~days is $\approx 8400$~km s$^{-1}$. The line profile is shown in Figure~\ref{fig:Fig6} with a clear double-peaked structure, possibly indicating a high-velocity, bi-polar outflow, or a toroidal geometry. It is possible though, that the redshifted peak is due instead to contamination from the [Si~\textsc{i}] feature at 1.099~$\mu$m.  The blueward peak also appears relatively weaker compared to the red side in the second epoch, and the full profile has narrowed to a FWHM velocity of $\approx 7600$~km~s$^{-1}$. The lower S/N of the second spectrum, however, adds difficulty in evaluating the significance of the profile evolution between the two epochs. In the FIRE spectrum, the narrow ($\mathrm{FWHM} \approx 300$~km~s$^{-1}$), redshifted dip in the profile is likely an artifact. This dip is only two pixels wide and is narrower than a resolution element ($R_{\mathrm{FIRE}} = 500$). Moreover, it is not detected in the higher resolution MOSFIRE spectrum ($R_{\mathrm{MOSFIRE}} = 3400$).

The flux in this line appears to have faded by a factor of more than 10 during the 17~days between the first and second epochs. Extrapolating from the observed $Y$-band decay rate of $0.020 \pm 0.005$~mag~day$^{-1}$ between 106 and 161 days, however, we would only expect the flux to have faded by a factor of $\approx 1.4$. We caution that the spectral flux calibrations using standard star observations are somewhat uncertain, and we do not have contemporaneous photometric data to verify the fading of this line. Slit losses due to a misalignment may also contribute to the apparent drop in flux of the He~\textsc{i} line rather than true variability in this feature. This could also explain the lack of detection of the continuum emission or any other features in the MOSFIRE spectrum.

The identification of the $1.083~\mu$m line is somewhat uncertain because we do not detect the He~\textsc{i} line at $2.058~\mu$m in either spectrum. It is possible that there is a contribution to the red wing of the $1.083~\mu$m feature from the Pa~$\gamma$ $1.094~\mu$m line, but we do not detect Pa~$\beta$ or Pa~$\delta$ ($1.282$ and $1.005~\mu$m, respectively), indicating that H~\textsc{i} is absent from the spectrum at this phase. We also note the $1.083~\mu$m He~\textsc{i} line profile may be contaminated by emission from the [Si~\textsc{i}] transitions at $1.082$ and $1.099~\mu$m.

Also shown in Figure~\ref{fig:Fig4} are the NIR spectra of the Type IIb SN~2011dh at phases of $\approx 170$~days, which show distinct similarity to those of SPIRITS\,15c. In addition to the broad, prominent He~\textsc{i} line at $1.083~\mu$m, several other lines identified in the spectrum of SN~2011dh are also present in the spectrum of SPIRITS\,15c. These include the Ca~\textsc{ii} IR triplet, the blended O~\textsc{i} and [Co~\textsc{ii}] feature near $0.93~\mu$m, the [Co~\textsc{ii}] feature near $1.02~\mu$m, the Mg~\textsc{i} line at $1.504~\mu$m, and the feature near $1.64~\mu$m due to either [Fe~\text{ii}] or [Si~\textsc{i}]. We discuss a detailed comparison between the NIR spectra of SPIRITS\,15c and SN~2011dh below in \S~\ref{sec:SN_spec}.

\section{Discussion} \label{sec:discussion}

Here we discuss possible physical interpretations of SPIRITS\,15c and SPIRITS\,14buu and compare them to other classes of transients discovered in recent years.

\subsection{SPIRITS\,15c as an obscured SN: comparison to SN~2011dh}\label{sec:obscured_SN}
We examine the possibility that SPIRITS\,15c is an SN explosion with significant obscuration by dust. We compare the properties of SPIRITS\,15c to the well studied Type~IIb explosion in M51, SN~2011dh, based on the similarity of their NIR spectra at late phases (\S~\ref{sec:SN_spec}). Throughout this section, following \citep{ergon14}, we adopt a distance to M51 of 7.8~Mpc, and a reddening along the line of sight to SN~2011dh of $E(B-V)=0.07$~mag assuming a standard $R_V = 3.1$ Milky Way extinction law as parametrized by \citep{fitzpatrick99}.

\subsubsection{NIR spectrum comparison}\label{sec:SN_spec}
The NIR spectrum of SPIRITS\,15c at $t=205$~d shows distinct similarity to the NIR spectra of SN~2011dh at similar phases. We show a direct comparison between the spectra of these objects in Figure~\ref{fig:Fig4}, where the phase of the SN~2011dh spectra is set so that the earliest detection of the SN coincides with the most constraining non-detection preceding the outburst of SPIRITS\,15c. The SN~2011dh spectra, originally published in \citet{ergon15}, were obtained from the Weizmann Interactive Supernova data REPository \citep[WISeREP,][]{yaron12}\footnote{WISeREP spectra are available here:\\\url{http://wiserep.weizmann.ac.il/}}. To aide in the comparison, we mark the features identified in the spectra of SN~2011dh by \citet{ergon15} and \citet{jerkstrand15} in the figure. 

A prominent feature in the spectra of both objects is the strong, broad ($\gtrsim 8000$~km~s$^{-1}$) He~\textsc{i} emission at $1.083~\mu$m, shown in detail in Figure~\ref{fig:Fig6}. This line is detected in both spectra of SPIRITS\,15c at $t=205$ and $222$~days. \citet{ergon15} note a blueshifted P-Cygni absorption component of this line profile in the SN~2011dh spectrum extending to a velocity of at least $\approx 10000$~km~s$^{-1}$ that may also be present in the line profile of SPIRITS\,15c at $t=202$~days. The He~\textsc{i} line profile of SPIRITS\,15c has a clear double-peaked structure that is not observed in SN~2011dh. This may suggest a bi-polar outflow or toroidal geometry in SPIRITS\,15c. We note that double-peaked [O~\textsc{i}] lines present in many stripped-envelope CCSNe have been interpreted as evidence for a toroidal or disk-like geometry \citep[e.g.][]{maeda02,mazzali05,maeda08,modjaz08,milisavljevic10}. The shoulder feature on the red side of the SN~2011dh line profile, likely due to contamination from the [Si~\textsc{i}] feature at $1.099~\mu$m, may contribute to the redshifted peak of the line profile in SPIRITS\,15c. 

The He~\textsc{i} line at $2.058~\mu$m, detected in the spectrum of SN~2011dh, is not detected in either spectrum of SPIRITS\,15c. Assuming the same flux ratio between the $1.083$ and $2.058~\mu$m He~\textsc{i} lines as is observed in SN~2011dh and accounting for additional reddening as inferred for SPIRITS\,15c below in \S~\ref{sec:SN_lc}, we would expect the $2.058~\mu$m He~\textsc{i} to peak at $0.09 \times 10^{-16}$~erg~s$^{-1}$~cm$^{-2}$~\AA$^{-1}$ above the continuum in the FIRE spectrum. This is below the RMS noise of $0.11$~erg~s$^{-1}$~cm$^{-2}$~\AA$^{-1}$ in this region of the spectrum, and thus, a non-detection of this feature is not surprising.

Other prominent emission features detected in the spectra of both objects include the Ca~\textsc{ii} IR triplet (0.845, 0.854, and 0.866~$\mu$m), the blended features due to O~\textsc{i} and [Co~\textsc{ii}] near $0.93~\mu$m, the [Co~\textsc{ii}] feature near $1.02~\mu$m, the Mg~\textsc{i} line at $1.504~\mu$m, and the feature near $1.64~\mu$m due to [Fe~\textsc{ii}] or [Si~\textsc{i}]. The excess emission beyond $2.3~\mu$m, attributed to the $\Delta v = 2$ vibrational overtones of CO by \citet{ergon15} and \citet{jerkstrand15} in the spectrum of SN~2011dh, appears even stronger in SPIRITS\,15c, but we note the spectrum of SPIRITS\,15c becomes increasingly noisy beyond $\approx 2.4~\mu$m at the end of $K$-band. Some of the weaker spectral features labeled in Figure~\ref{fig:Fig4} and present in the spectrum of SN~2011dh are not detected in SPIRITS\,15c, but this can likely be accounted for by the relatively lower S/N of the SPIRITS\,15c spectra, small intrinsic differences in the strength of these features between the two events, and the uncertainty of matching the evolutionary phase between the two events. Overall, we find that the late-time, NIR spectrum of SN~2011dh provides a good match to the spectra of SPIRITS\,15c, and we evaluate the interpretation of SPIRITS\,15c as a Type~IIb SN similar to SN~2011dh, though subject to significant dust obscuration, below in Sections~\ref{sec:SN_prog} and \ref{sec:SN_lc}.

\subsubsection{Progenitor comparison}\label{sec:SN_prog}

As discussed above in \S~\ref{sec:hst}, no progenitor star was detected at the position of SPIRITS\,15c in pre-explosion \textit{HST} imaging of IC~2163 and NGC~2207 to limiting magnitudes of $M_V > -7.9$ and $M_I > -8.9$~mag at the assumed distance to the host, correcting only for Galactic extinction. A candidate progenitor of SN~2011dh was identified as an intermediate-mass yellow supergiant star in pre-explosion \textit{HST} images at the position of the SN \citep{maund11,vandyk11}. The disappearance of this source in post-explosion imaging confirms it was indeed the progenitor star \citep{vandyk13,ergon14}. The observed magnitudes in the HST frames were $V \sim 21.8$ and $I \sim 21.2$, corresponding to absolute magnitudes of $M_V \sim -7.9$ and $M_I \sim -8.4$ at the assumed distance and reddening to SN~2011dh \citep{maund11, vandyk11}. These values are consistent with the non-detection of a progenitor star for SPIRITS\,15c in the archival \textit{HST} imaging of IC~2163, and furthermore, there may be significant, additional extinction from the foreground spiral arm of NGC~2207 and the local environment of the transient. Thus, we cannot rule out a progenitor system similar to that of SN~2011dh for SPIRITS\,15c.

\subsubsection{Light curve and SED comparison}\label{sec:SN_lc}
The optical and IR light curves of SPIRITS\,15c are shown in Figure~\ref{fig:Fig2}. The peak observed brightness of the transient in the optical was in $i$-band at $i=17.67\pm0.09$~mag ($M_i = -15.08$~mag absolute). The peak in $i$-band absolute magnitude of SN~2011dh was brighter by $\approx 2$~mag, possibly indicating that SPIRITS\,15c is subject to significant dust extinction. To examine this scenario, we compare the light curves of SN~2011dh from \citet{helou13}, \citet{ergon14}, and \citet{ergon15} to those of SPIRITS\,15c in Figure~\ref{fig:Fig2}. The light curves of SN~2011dh have been shifted in apparent magnitude to the assumed distance of SPIRITS\,15c, and are shifted in phase so that the earliest detection of SN~2011dh coincides with the last non-detection of SPIRITS\,15c before its outburst. We then applied a \citet{fitzpatrick99} extinction law with $R_V = 3.1$ and $E(B-V) = 0.72$~mag, corresponding to $A_V = 2.2$~mag, to the SN~2011dh light curves. As shown in the figure, this provides a good match to the observed magnitudes of SPIRITS\,15c in $gri$ at $t=38.6$~days. We note a few small discrepancies with the optical light curves in this comparison. Namely, SPIRITS\,15c is fainter than the adjusted brightness of SN~2011dh at $t=0$~days in $i$-band by $\approx 0.3$~mag and brighter by $\approx 0.1$~mag at $t=30$~days.

While the reddened light curves of SN~2011dh provide a good match to the optical light curves of SPIRITS\,15c, the IR light curves are more difficult to explain using only a standard extinction law. For the extinction and phase assumed for SN~2011dh in Figure~\ref{fig:Fig2}, several discrepancies are apparent. At $t=106$~days, the $H$-band magnitude is brighter than the corresponding value for SN~2011dh by $0.6$~mag. Furthermore, the SN~2011dh $H$-band light curve decays faster, resulting in a larger discrepancy of $1.0$~mag by $t=161$~days. At the same phase, the adjusted magnitude of SN~2011dh is $0.9$~mag fainter than that of SPIRITS\,15c in $J$. The single $K$-band detection of SPIRITS\,15c at $181$~days matches the reddened light curve of SN~2011dh quite well.

The largest discrepancy occurs in the \textit{Spitzer}/IRAC [3.6] and [4.5] bands. In [4.5], the SPIRITS\,15c light curve is brighter the than that of SN~2011dh by $\approx 0.4$~mag between $t=167$ and $307$~days, though they decay at similar rates. However, in [3.6] at $t=167$~days, SPIRITS\,15c is fainter than the reddened SN~2011dh by $1.5$~mag. To explain the $1.9$~mag excess in $[3.6]-[4.5]$ color of SPIRITS\,15c over SN~2011dh using only a standard reddening law would require $\gtrsim 100$~magnitudes of visual extinction.

\begin{figure}
\centering
\includegraphics[width=\linewidth]{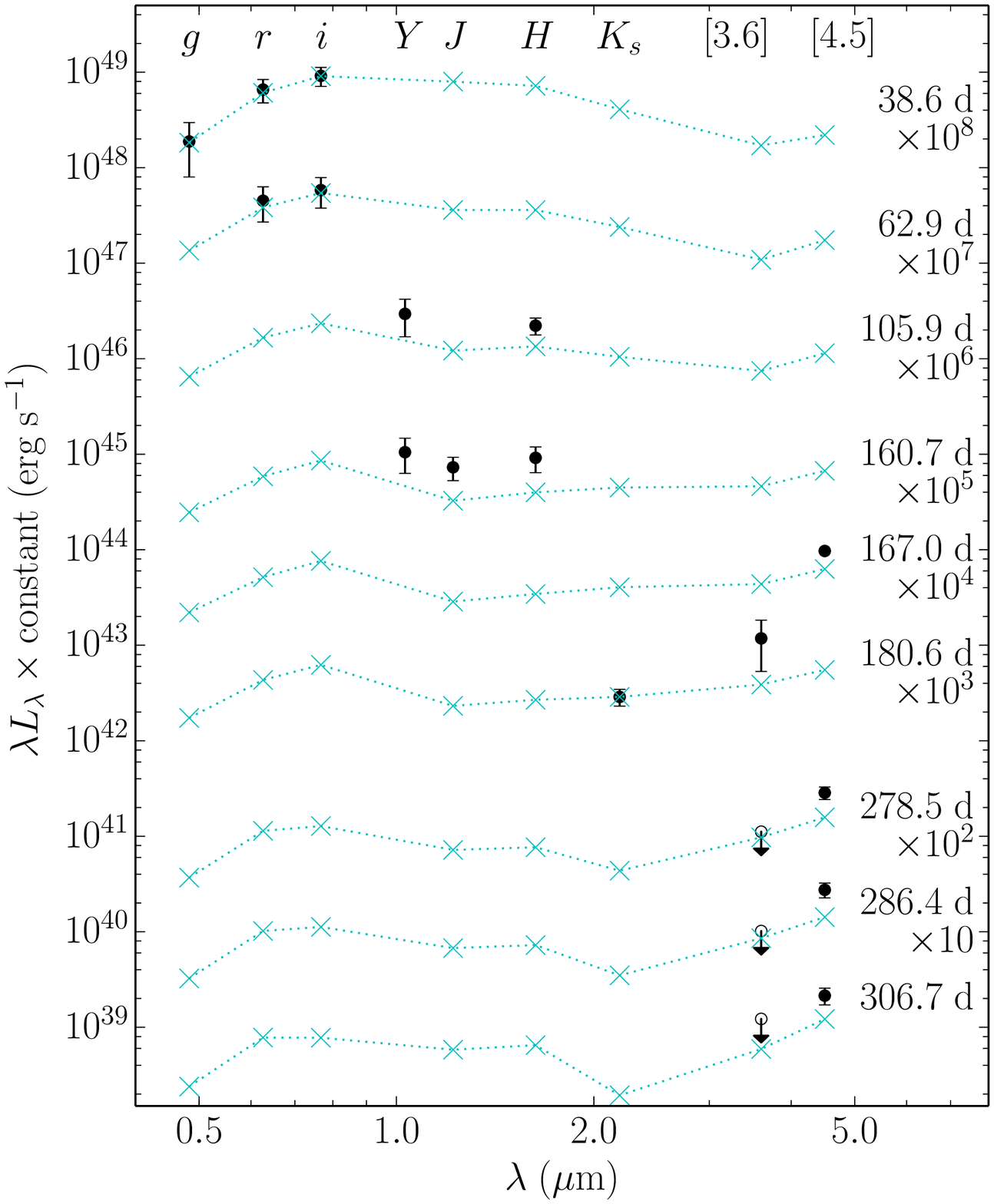}
\caption
{ \label{fig:Fig7}
Comparison of the SED evolution of SPIRITS\,15c (black circles) and SN~2011dh (cyan `X'-symbols). Unfilled points with downward arrows represent upper-limits. The band-luminosities for SPIRITS\,15c are corrected for Galactic extinction to IC~2163 from NED \citep{fitzpatrick99, schlafly11, chapman09}. The points for are SN~2011dh drawn from a linear interpolation of the SN~2011dh light curves, and reddened by the assumed excess extinction in SPIRITS\,15c characterized by $E(B-V)_{\rm host} = 0.65$ with a standard $R_V = 3.1$ extinction law. Error bars for SPIRITS\,15c are shown, but are sometimes smaller than the plotting symbols. The phase, set as in Figure~\ref{fig:Fig2}, of each SED is listed along the right side, and each SED is shifted up by a factor of 10 from the one below for clarity.}
\end{figure}

We make a similar comparison between the SED of SPIRITS\,15c and the reddened SED of SN~2011dh at several phases in Figure~\ref{fig:Fig7}. As in the light curve comparison, we find that applying extinction with $E(B-V) = 0.72$ to the SED of SN~2011dh at $t=39$ and $63$~days does a good job reproducing the observed optical $gri$ band-luminosities of SPIRITS\,15c. The IR $YJH$, and [4.5] band-luminosities, however are significantly over-luminous compared to the reddened SN~2011dh between $t=106$ and $307$~days. Again, the most apparent discrepancy in this comparison is the significant over-prediction of the luminosity of SPIRITS\,15c at [3.6].

Because SPIRITS\,15c appears over-luminous in some bands, and under-luminous in others compared to the reddened light curves and SED of SN~2011dh, a phase shift in the light curves of SN~2011dh cannot account for the observed discrepancies. Furthermore, using a steeper or shallower extinction law to redden the observations of SN~2011dh would be unable to rectify all of the observed discrepancies at once.

\subsubsection{SPIRITS\,15c in the context of SNe IIb}
The designation IIb indicates an SN that transitions spectroscopically from a hydrogen-rich SN~II to a hydrogen-poor SN~Ib. These events are believed to arise from the core collapse of massive stars that have been stripped of almost all of their hydrogen envelope. In addition to SN~2011dh, well-studied examples of SNe~IIb include SN~1993J, SN~2008ax. 

The progenitor of SN~1993J was shown to be a yellow, extended supergiant star of 12--17~$M_{\odot}$, similar to that of SN~2011dh, and it was proposed to have lost most of its hydrogen envelope to a blue, compact companion star \citep{podsiadlowski93,shigeyama94,woosley94,blinnikov98,maund04,stancliffe09}. The progenitor observations of SN~2008ax by \citet{crockett08} were less conclusive, but \citet{tsvetkov09} proposed a $13~M_{\odot}$ star with an extended, low-mass hydrogen envelope based on hydrodynamical modeling of the SN optical light curves. As mentioned above in \S~\ref{sec:SN_prog}, we do not rule out a progenitor similar to those of other SNe~IIb based on pre-explosion HST imaging. 

\citet{ergon14} showed that these SNe had similar optical light curve and color evolution in the first 50 days. All three reached peak luminosity at $\approx 20$~days post explosion, and evolved to redder colors with time. They noted some differences between the three events, such as a factor of 3 spread in peak luminosity and a $\approx 0.4$~mag spread in their optical colors, but these differences could be accounted for within the systematic uncertainties in the distance and extinction to each SN. The assumption of $E(B-V)_{\rm host} = 0.65$~mag for SPIRITS\,15c brings its optical light curves and color evolution into good agreement with the observed properties of well-studied SNe~IIb. 

The IR light curves of SPIRITS\,15c are not as well matched by a reddened version of SN~2011dh, but it is not as straight forward to draw a direct comparison between the two events at these wavelengths. SN~2011dh was shown to have a significant IR excess at [4.5] compared to the NIR and [3.6] measurements, possibly due to a thermal echo from heated CSM dust, dust formation in the ejecta, and/or CO fundamental band emission \citep{helou13,ergon14, ergon15}. Similar confounding factors may be at play in SPIRITS\,15c given the observed excess at NIR wavelengths and at [4.5], and the possible CO $\Delta v = 2$ vibrational emission present in the K-band spectrum. The extreme $[3.6] - [4.5] = 3.0 \pm 0.2$~mag color of SPIRITS\,15c at a phase of 167~days has not been observed for another SN~IIb, and provides the strongest challenge to the interpretation of SPIRITS\,15c as a moderately obscured SN~2011dh-like event. SN~2011dh only reached a \textit{Spitzer}/IRAC color of 1.6~mag at $\approx 400$~days, and the Type~IIb SN~2013df was observed to have a $1.3$~mag color at $\approx 300$~days \citep{tinyanont16}.

Although SPIRITS\,15c and SN~2011dh show remarkable similarity in many respects, we note that we do not have sufficient data discriminate between a SN~IIb and SN~Ib in this case. As is clearly demonstrated by \citet{fremling16}, the light curves of PTF\,12os (Type~IIb SN~2012P) and PTF\,13bvn (Type~Ib SN) show only very minor differences to each other and to SN~2011dh. Spectroscopically, SNe~Ib and IIb are only distinguished by the presence or absence of hydrogen in their early-time spectra. The absence of hydrogen in the spectrum of SPIRITS\,15c after $\approx 200$~days does not classify this event as a SN~Ib, as hydrogen may very well have been present at earlier times. 

\subsection{SPIRITS\,14buu: Yet another reddened SN?}\label{sec:IC2163OT_SN}
The optical light plateau and subsequent fall-off observed for SPIRITS\,14buu are similar to those characteristic of the SN~IIP class of CCSNe. The duration of SNe~IIP plateaus is typically $\approx 100$~days \citep[e.g][]{poznanski09,arcavi12}. \citet{sanders15} found using a statistical sample of SN~IIP light curves from Pan-STARRS1 a median $r$-band plateau duration of 92 days with a 1$\sigma$ variation of 14 days, though the full distribution spanned a range of 42--126 days. The observed $>80$~day plateau of SPIRITS\,14buu in the $r$ and $i$ bands is thus consistent with the observed distribution for SNe~IIP. 

SNe~IIP have been observed to exhibit a wide-ranging continuum in observed peak magnitude in the optical ($-14.5 \gtrsim M_r \gtrsim -20.0$, $-15.0 \gtrsim M_i \gtrsim -19.5$, \citealp{sanders15}), with higher luminosity events exhibiting greater expansion velocities and producing larger amounts of nickel \citep[e.g.][]{hamuy03}. Correcting only for galactic extinction, SPIRITS\,14buu is fainter than the faintest SN~IIP in the \citet{sanders15} sample by $\approx 1.5$~mag in both the $r$ and $i$ bands. It is possible IC~2163 is an intrinsically low-luminosity SN~IIP, however the observed red colors may also indicate additional dust extinction. 

To examine this possibility we compare SPIRITS\,14buu to the low-luminosity Type IIP SN~2005cs in M51 in Figure~\ref{fig:Fig3}. We obtained $BVRIJH$ light curves of SN~2005cs from and assume extinction with $E(B-V) = 0.05$~mag following \citet{pastorello09}. For a direct comparison, the optical measurements were converted to SDSS system magnitudes using the conversions from \citep{jordi06}, and corrected for Galactic extinction to M51. Following a similar method to that used in \S~\ref{sec:SN_lc}, we then shifted the SN~2005cs light curves in apparent magnitude to the distance of IC\,2163, and applied a total reddening characterized by $E(B-V) = 0.49$~mag with a standard $R_V = 3.1$ \citet{fitzpatrick99} extinction law, corresponding to $1.5$~magnitudes of visual extinction. We also shifted SN~2005cs light curves by an additional $0.1$~mag (smaller than the uncertainties in the distance moduli to the hosts) to achieve a better match to the $r$ and $H$-band measurements for SPIRITS\,14buu. The phase of SN~2005cs was set so that the fall-off of the optical plateau coincides with that of SPIRITS\,14buu. 

The reddened light curves of SN~2005cs provide a good match to the $r$, $J$, and $H$-band light curves of SPIRITS\,14buu. The $g$-band upper limits for SPIRITS\,14buu are also mostly consistent with the reddened $g$-band light curve of SN~2005cs. The largest discrepancy is in the $i$-band between 60--100 days, where SPIRITS\,14buu is $0.5$~mag fainter than the reddened SN~2005cs. 

We also compare the [3.6] and [4.5] light curves of SPIRITS\,14buu to those of the Type IIP SN~2004et from \citet{kotak09} in Figure~\ref{fig:Fig3}. The SN~2004et light curves are shifted fainter in absolute magnitude by 2.4~mag to match the level of SPIRITS\,14buu. This also suggests SPIRITS\,14buu may be an intrinsically faint event like SN~2005cs, in addition to suffering from significant extinction. The decay rate of SN~2004et is somewhat faster than SPIRITS\,14buu in both \textit{Spitzer}/IRAC bands. \citet{tinyanont16} present a sample of SN light curves at [3.6] and [4.5] and found that SN~II mid-IR light curves show a large degree of variety. Thus we would not necessarily expect to find a good match to the mid-IR of SPIRITS\,14buu in the relatively small sample of SNe~IIP that have been well studied at these wavelengths.

Without a spectrum, we cannot definitively determine the nature of this source. Overall, however, we find a low-luminosity SN~IIP, similar to SN~2005cs with $\approx 1.5$~mag of visual extinction, to be a reasonable explanation for the observed properties of SPIRITS\,14buu. 

\subsection{Non-supernova IR transient scenarios}\label{sec:non-SN}
We also consider the possibility that SPIRITS\,15c and SPIRITS\,14buu are not extinguished SNe. Below, we find that scenarios involving a stellar merger and an SN~2008S-like transient can likely be ruled out for both events. We also consider a V445~Pup-like helium nova, which can likely be ruled out in the case of SPIRITS\,15c due to its high luminosity. 

\subsubsection{Stellar merger}
Known transients associated with stellar merger events include the low-mass (1 -- 2~$M_{\odot}$) contact binary merger V1309 Sco \citep{tylenda11}, and the more massive merger of a B-type progenitor (5 -- 10~$M_{\odot}$) V838 Mon \citep{bond03, sparks08}. Recently, the NGC~4490-OT has also been proposed to be a high-mass (20 -- 30~${M_{\odot}}$) analog of these events \citep{smith16}. The observed examples of stellar mergers are characterized by unobscured, optical progenitors, irregular multi-peaked light curves, increasing red colors with time, and a significant IR excess at late times. The spectra during the decline phase show relatively narrow lines ($O[10^2]$~km~s$^{-1}$) of H~\textsc{i} in emission and Ca~\textsc{ii} in absorption.

The light curve of SPIRITS\,15c shows some similarities to the massive stellar merger NGC4490-OT although it is $\approx\ $1--2~magnitudes brighter both during the optical peak, and in the IR \textit{Spitzer}/IRAC bands at late times. The $[3.6] - [4.5]$ color of SPIRITS\,15c is significantly redder. The optical outburst of the NGC~4490-OT lasted longer than 200~days and the IR light curve extends in excess of 800~days. SPIRITS\,15c, in contrast, faded beyond detection in the optical in $\lesssim 150$~days and in the IR in $\lesssim 500$~days. In the case of the NGC~4490-OT, the evidence for a massive stellar merger is corroborated by the detection of a massive progenitor at $M_{F606W}\approx-6.4$, likely an $M \approx 30~M_{\odot}$ LBV. This extends the observed correlation suggested by \citet{kochanek14} in peak luminosity and merger progenitor mass \citep{smith16}. Further extending this correlation to the luminosity of SPIRITS\,15c would suggest an extremely massive progenitor ($M \gtrsim 60~M_{\odot}$). The progenitor of SPIRITS\,15c was undetected in the 1998 HST images to $M_V > -7.9$~mag. Furthermore, the velocities observed in SPIRITS\,15c are $\gtrsim 10$ times those seen in stellar mergers, indicating a significantly more explosive event. Finally, the lack of hydrogen observed for SPIRITS\,15c would require a rare, very massive stripped-envelope system. Taken together, the evidence indicates that a massive stellar merger scenario is likely ruled out as an explanation for the properties of SPIRITS\,15c.

SPIRITS\,14buu reached a similar peak luminosity in the $i$-band to the peak optical luminosity of the NGC~4490-OT, but SPIRITS\,14buu is $\approx 1$~mag brighter at [4.5]. As with SPIRITS\,15c, the NGC~4490-OT evolved much more slowly than SPIRITS\,14buu in both the optical and IR. Furthermore, the observed plateau of SPIRITS\,14buu in the optical is not consistent with the characteristic multi-peaked light curves of stellar merger events. We find that despite the similarity in observed optical luminosity, SPIRITS\,14buu is overall inconsistant with an NGC~4490-OT-like stellar merger. 

\subsubsection{SN~2008S-like transient}
Supernova~(SN)~2008S-like transients, including the 2008 luminous optical transient in NGC~300 (NGC~300~OT2008-1) and other similar events, are a class of so-called SN imposters. These events have highly obscured progenitors in the optical, but bright mid-IR (MIR) detections ($M_{[4.5]} < -10$) suggest that the progenitors are likely extreme asymptotic giant branch stars (EAGB) of intermediate mass ($\approx 10~\mathrm{to}~15~M_{\odot}$), self-obscured by a dense wind of gas and dust \citep{prieto08,bond09,thompson09}. They are significantly less luminous than SNe in the optical, with peak absolute visual magnitude of only $M_V \approx -13$ for SN~2008S \citep{steele08} and $M_V \approx -12~\mathrm{to}~-13$ for NGC~300~OT2008-1 \citep{bond09}. These events have peculiar spectral features including narrow emission from [Ca~\textsc{ii}], Ca~\textsc{ii}, and weak Fe~\textsc{ii} \citep{steele08, bond09, humphreys11}. Moderate resolution spectroscopy of NGC~300~OT2008-1 revealed that the emission lines showed a double peaked structure, indicating a bipolar outflow with expansion velocities of $\approx 70 - 80$~km~s$^{-1}$ \citep{bond09, humphreys11}.  A proposed physical picture of these events is that a stellar explosion or massive eruption, possibly an electron-capture SN, destroys most of the obscuring dust, allowing the transient to be optically luminous. In the aftermath, however, the dust reforms and re-obscures the optical transient, producing a significant IR-excess \citep{thompson09,kochanek11}. Both transients have now faded beyond their progenitor luminosities at [3.6] and [4.5], suggesting the outbursts were terminal events \citep{adams16}. 

The optical and late-time MIR luminosities of SPIRITS\,15c are 1--2 magnitudes brighter than is observed for SN~2008S-like events, and SPIRITS\,15c is again significantly redder in the MIR (only $\approx 1$~mag for SN~2008S-like events, \citealp{szczygiel12}). As with SN~2008S and the NGC~300~OT2008-1, the progenitor of SPIRITS\,15c was obscured in the optical. There was no detection of a MIR progenitor for SPIRITS\,15c down to $\approx 14.9$~mag ($\approx -17.85$ absolute), but this limit is not constraining due to the larger distance to IC~2163 and the bright background galaxy light in the vincity of SPIRITS\,15c. The observed velocities of SPIRITS\,15c are much higher than those of SN~2008S or the NGC~300~OT2008-1, which show similar velocities to stellar mergers. The strongest evidence against SPIRITS\,15c as a SN~2008S-like event is the lack of hydrogen in its spectrum. SN~2008S and the NGC~300~OT2008-1 both showed strong hydrogen emission features, and moreover, if the physical mechanism behind them is indeed an electron-capture SN, the lack of hydrogen in SPIRITS\,15c would be inexplicable given the believed intermediate mass ($M \approx 8 - 10~M_{\odot}$) progenitors of such events. 

The $i$-band luminosity of SPIRITS\,14buu is comparable to SN~2008S-like event in the optical, and both SPIRITS\,14buu and SN~2008S reached IR luminosities of $M_{[4.5]} \approx -16$. SN~2008S was also significantly redder in $[3.6] - [4.5]$ color by $\approx 0.8$~mag. Additionally, SN~2008S was fainter than SPIRITS\,14buu in the NIR by $\approx 2$~magnitudes. The largest discrepancy is that the optical light curves of SN~2008S decayed steeply after only $\approx 30$~days in constrast with the $\gtrsim 80$~day plateau of SPIRITS\,14buu. Thus, we find a SN~2008S-like events to be a poor match to the observed properties of SPIRITS\,14buu.  

\subsubsection{V445 Pup-like transient or a helium nova}
An intriguing possibility for an explanation of SPIRITS\,15c is a helium nova, a rare and so far largely theoretical class of events. Helium novae are due to thermonuclear runaway on the surface of a white dwarf accreting helium from a helium-rich donor \citep{kato89}. The most convincing and well studied candidate for an observed helium nova is the 2000 outburst of V445~Pup. The optical outburst showed a similar decline to slow novae, but was small in amplitude with $\Delta m_V \approx 6$~mag. The spectrum was also unusual for a classical nova in that it was rich in He~\textsc{i} and C~\textsc{i} emission lines, but showed no evidence for hydrogen. Multi-epoch adaptive optics imaging and integral field unit spectroscopy revealed an expanding bipolar outflow from V445~Pup with a velocity of $\approx 6700$~km~s$^{-1}$ and some knots at even larger velocities of $\approx 8500$~km~s$^{-1}$ \citep{woudt09}. Using their observations of the expanding shell, \citet{woudt09} derive an expansion parallax distance of $8.2 \pm 0.5$~kpc. 

The broad, double-peaked He~\textsc{i} line in the spectrum of SPIRITS\,15c is in good agreement with the observed velocities of the bipolar helium outflow of V445~Pup. The largest problem with this scenario, however, is that the V445~Pup outburst was significantly less luminous than SPIRITS\,15c in the optical, with a peak of only $M_V \approx -6$. The NIR of peak at $M_{K_s} \approx -10$ was more comparable to that of SPIRITS\,15c, and both sources grew redder with time during the initial decline. Late-time observations of V445~Pup with \textit{Spitzer}/IRAC in 2005 (P20100; PIs Banerjee \& P. Dipankar) and 2010 (P61071; PI B.~A. Whitney), approximately $5~\mathrm{and}~10$~years after the outburst, respectively, revealed the source was slowing decaying at [3.6] from $m_{[3.6]} \approx 8.6~\mathrm{to}~9.5$ ($-6~\mathrm{to}~-5.1$ absolute) over $\approx 5$~years. The source was saturated in the IRAC2, IRAC3, and IRAC4 channels indicating very red lower limits on the MIR colors of $[3.6]-[4.5] \gtrsim 2.3$~mag, $[3.6]-[5.7] \gtrsim 3.4$~mag, and  $[3.6]-[7.9] \gtrsim 4.2$~mag. 

\section{Conclusions} \label{sec:conclusions}
SPIRITS\,15c is a luminous ($M_{[4.5]} = -17.1 \pm 0.4$), red ($[3.6] - [4.5] = 3.0 \pm 0.2$~mag), IR-dominated transient discovered by the SPIRITS team. The transient was accompanied by an optical precursor outburst with $M_i = -15.1 \pm 0.4$~mag that was quickly overtaken by IR emission within $\approx 100$~days. The most prominent feature of the NIR spectrum is a broad ($\approx 8000$~km~s$^{-1}$), double-peaked emission line at $1.083~\mu$m, likely due to He~\textsc{i}, and possibly indicating a He-rich, bi-polar or toroidal outflow associated with the transient. 

We explored several possible scenarios to explain the unusual observed properties of SPIRITS\,15c. Both the stellar merger and SN~2008S-like scenarios can likely be ruled out by the high luminosity of SPIRITS\,15c and the explosive velocities and lack of hydrogen in its spectrum. In the case of a helium nova, the strong He~\textsc{i} emission and high-velocity, bipolar outflow of SPIRITS\,15c is similar to that observed in the candidate prototype helium nova V445 Pup, but again, the optical luminosity of SPIRITS\,15c is too extreme. 

We conclude that SPIRITS\,15c is a stripped envelope, CCSN explosion similar to the well studied Type~IIb SN~2011dh, but extinguished by dust in the optical and NIR at a level of $A_V \approx 2.2$~mag. The spectrum of SPIRITS\,15c is very similar to that of SN~2011dh at a phase of $\approx 200$~days, but SPIRITS\,15c shows a distinct double-peaked profile in the broad, strong He~\textsc{i} emission line that is not observed in SN~2011dh. The assumption of $A_V = 2.2$~mag with a standard $R_V = 3.1$ extinction law in SPIRITS\,15c provides a good match between the optical light curves and observed colors of SPIRITS\,15c and those of SN~2011dh. In the IR, however, SPIRITS\,15c is more luminous than SN~2011dh, except at [3.6] where it is significantly under-luminous, illustrated by its extreme $[3.6]-[4.5]$ color. Thus, we find that a shift in phase or simply a steeper extinction law cannot explain the observed differences between SPIRITS\,15c and SN~2011dh. 

SPIRITS\,14buu, an earlier transient in IC\,2163, was serendipitously discovered in the SPIRITS data during the analysis of SPIRITS\,15c. The source was also luminous in the IR at $M_{[4.5]} = -16.1 \pm 0.4$~mag, and developed a fairly red $[3.6] - [4.5]$ color of $1.2 \pm 0.2$~mag by 166~days after its first detection. The optical and NIR light curves showed plateau lasting at least 80 days, similar to that observed in SNe~IIP. Scenarios involving a stellar merger or SN~2008S-like transient can again likely be ruled out. A comparison to the low-luminosity Type IIP SN~2005cs assuming $\approx 1.5$~magnitudes of visual extinction produced a reasonable match to the properties of SPIRITS\,14buu, and we find an obscured SN~IIP to be the most likely interpretation for SPIRITS\,14buu considered here. 

The key to fully understanding the nature of these events is MIR spectroscopy, the likes of which will become available with the launch of the James Webb Space Telescope (JWST). A MIR spectrum would, for example, enable us to elucidate the origin of the extreme $[3.6]-[4.5]$ color observed in SPIRITS\,15c, and identify spectral features contributing to the flux at these wavelengths, e.g., the fundamental vibrational overtones of CO that may contribute to the high luminosity at [4.5]. Moreover, photometric coverage further into the MIR with JWST would allow us to detect the presence of a cooler dust component than is accessible with the warm \textit{Spitzer}/IRAC bands.

The census of SNe in nearby galaxies from optical searches, even at only 35~Mpc, is incomplete. In less than 3 years of monitoring nearby galaxies at IR wavebands, SPIRITS has discovered at least one, and possibly two, moderately extinguished SNe that went unnoticed by optical surveys, SPIRITS\,15c and SPIRITS\,14buu. Since the start of the SPIRITS program, our galaxy sample has hosted 9 optically discovered CCSNe. This may suggest a rate of CCSNe in nearby galaxies missed by current optical surveys of at least $10\%$. If SPIRITS\,14buu is also indeed an SN, this estimate increases to 18\%. In a future publication, we will present an analysis of the full SPIRITS sample of SN candidates to provide a more robust estimate of the fraction of SNe being missed by optical surveys. If this fraction is high, this could have significant implications for our understanding of the CCSN rate and its connection to star formation rates. Additionally, the discovery of IR transients such as SPIRITS\,15c and SPIRITS\,14buu in a galaxy-targeted survey indicates that the night sky is ripe for exploration by dedicated wide-field, synoptic surveys in the IR.

\begin{acknowledgements}
We thank M. Kaufman, B. Elmegreen, and the authors of \citet{elmegreen16} for providing H$\alpha$, $24~\mu\mathrm{m}$, and CO measurements at the location of SPIRITS\,15c. We also thank P. Groot, R. Lau, and S. Adams for valuable discussions.

This work is based in part on observations made with the Spitzer Space Telescope, which is operated by the Jet Propulsion Laboratory, California Institute of Technology under a contract with NASA. The work is based, in part, on observations made with the Nordic Optical Telescope, operated by the Nordic Optical Telescope Scientific Association at the Observatorio del Roque de los Muchachos, La Palma, Spain, of the Instituto de Astrofisica de Canarias.

This material is based upon work supported by the National Science Foundation Graduate Research Fellowship under Grant No. DGE-1144469. HEB acknowledges support for this work provided by NASA through grants GO-13935 and GO-14258 from the Space Telescope Science Institute, which is operated by AURA, Inc., under NASA contract NAS 5-26555. RDG was supported in part by the United States Air Force. 

\end{acknowledgements}



\onecolumngrid
\LongTables
\clearpage
\begin{deluxetable*}{lcccccc}
\tablecaption{Photometry of SPIRITS\,15c \label{table:phot}}
\tablehead{\colhead{UT Date} & \colhead{MJD} & \colhead{Phase\tablenotemark{a}} & \colhead{Tel./Inst.} & \colhead{Band} & \colhead{Apparent Magnitude\tablenotemark{b,c}} & \colhead{Absolute Magnitude\tablenotemark{c,d}} \\ 
\colhead{} & \colhead{} & \colhead{(days)} & \colhead{} & \colhead{} & \colhead{(mag)} & \colhead{(mag)} } 
\startdata
2013 Dec 16.1 & 56642.1 & -248 & NOTCam & $K_s$ & $>17.0$ & $>-15.8$ \\
2013 Dec 24.3 & 56650.3 & -240 & P60 & $g$ & $>20.4$ & $>-12.7$ \\
2014 Jan 13.9 & 56670.9 & -220 & \textit{Spitzer}/IRAC & $[3.6]$ & $>18.6$ & $>-14.2$ \\
2014 Jan 13.9 & 56670.9 & -220 & \textit{Spitzer}/IRAC & $[4.5]$ & $>18.4$ & $>-14.3$ \\
2014 Jan 10.2 & 56667.2 & -223 & du Pont/RetroCam & $Y$ & $>19.7$ & $>-13.2$ \\
2014 Jan 10.2 & 56667.2 & -223 & du Pont/RetroCam & $J$ & $>19.2$ & $>-13.6$ \\
2014 Jan 10.2 & 56667.2 & -223 & du Pont/RetroCam & $H$ & $>18.8$ & $>-14.0$ \\
2014 Feb 06.2 & 56694.2 & -196 & Baade/IMACS & $\text{WB6226-7171}$ & $>22.0$ & $>-10.9$ \\
2014 Feb 18.2 & 56706.2 & -184 & Swopes/CCD & $g$ & $>20.7$ & $>-12.3$ \\
2014 Feb 18.2 & 56706.2 & -184 & Swopes/CCD & $r$ & $>20.7$ & $>-12.2$ \\
2014 Feb 18.2 & 56706.2 & -184 & Swopes/CCD & $i$ & $>20.5$ & $>-12.4$ \\
2014 Mar 07.1 & 56723.1 & -167 & Baade/FourStar & $J$ & $>20.4$ & $>-12.5$ \\
2014 Mar 15.1 & 56731.1 & -159 & du Pont/RetroCam & $Y$ & $>20.7$ & $>-12.2$ \\
2014 Mar 15.1 & 56731.1 & -159 & du Pont/RetroCam & $J$ & $>20.4$ & $>-12.4$ \\
2014 Mar 16.1 & 56732.1 & -158 & du Pont/RetroCam & $H$ & $>19.0$ & $>-13.8$ \\
2014 Apr 13.0 & 56760.0 & -130 & Swopes/CCD & $r$ & $>20.6$ & $>-12.3$ \\
2014 Apr 13.0 & 56760.0 & -130 & Swopes/CCD & $i$ & $>20.5$ & $>-12.4$ \\
2014 May 13.7 & 56790.7 & -100 & \textit{Spitzer}/IRAC & $[3.6]$ & $>18.8$ & $>-14.0$ \\
2014 May 13.7 & 56790.7 & -100 & \textit{Spitzer}/IRAC & $[4.5]$ & $>18.6$ & $>-14.2$ \\
2014 May 27.9 & 56804.9 & -86 & LCOGT-1m/CCD & $i$ & $>18.7$ & $>-14.2$ \\
2014 Jun 08.7 & 56816.7 & -74 & \textit{Spitzer}/IRAC & $[3.6]$ & $>18.7$ & $>-14.0$ \\
2014 Jun 08.7 & 56816.7 & -74 & \textit{Spitzer}/IRAC & $[4.5]$ & $>18.3$ & $>-14.4$ \\
2014 Jul 25.4 & 56863.4 & -27 & LCOGT-1m/CCD & $i$ & $>19.0$ & $>-13.9$ \\
2014 Aug 21.4 & 56890.4 & 0 & LCOGT-1m/CCD & $i$ & $17.66$ $(0.06)$ & $-15.2$ \\
2014 Sep 20.3 & 56920.3 & 30 & LCOGT-1m/CCD & $i$ & $18.25$ $(0.06)$ & $-14.6$ \\
2014 Sep 29.0 & 56929.0 & 39 & Swopes/CCD & $g$ & $20.9$ $(0.2)$ & $-12.1$ \\
2014 Sep 29.0 & 56929.0 & 39 & Swopes/CCD & $r$ & $19.2$ $(0.1)$ & $-13.7$ \\
2014 Sep 29.0 & 56929.0 & 39 & Swopes/CCD & $i$ & $18.57$ $(0.09)$ & $-14.3$ \\
2014 Oct 23.3 & 56953.3 & 63 & LCOGT-1m/CCD & $r$ & $19.6$ $(0.3)$ & $-13.3$ \\
2014 Oct 23.3 & 56953.3 & 63 & LCOGT-1m/CCD & $i$ & $19.1$ $(0.3)$ & $-13.8$ \\
2014 Nov 15.2 & 56976.2 & 86 & LCOGT-1m/CCD & $r$ & $20.4$ $(0.3)$ & $-12.6$ \\
2014 Nov 15.2 & 56976.2 & 86 & LCOGT-1m/CCD & $i$ & $19.5$ $(0.3)$ & $-13.4$ \\
2014 Nov 20.4 & 56981.4 & 91 & Lemmon-1.5m/2MASSCam & $J$ & $>14.3$ & $>-18.5$ \\
2014 Nov 20.4 & 56981.4 & 91 & Lemmon-1.5m/2MASSCam & $H$ & $>13.5$ & $>-19.3$ \\
2014 Nov 20.4 & 56981.4 & 91 & Lemmon-1.5m/2MASSCam & $K_s$ & $>13.1$ & $>-19.7$ \\
2014 Dec 05.3 & 56996.3 & 106 & du Pont/RetroCam & $Y$ & $18.8$ $(0.2)$ & $-14.1$ \\
2014 Dec 05.3 & 56996.3 & 106 & du Pont/RetroCam & $H$ & $17.79$ $(0.08)$ & $-15.0$ \\
2014 Dec 14.3 & 57005.3 & 115 & LCOGT-1m/CCD & $r$ & $>19.6$ & $>-13.3$ \\
2014 Dec 14.3 & 57005.3 & 115 & LCOGT-1m/CCD & $i$ & $>19.2$ & $>-13.7$ \\
2014 Dec 20.2 & 57011.2 & 121 & Swopes/CCD & $g$ & $>21.1$ & $>-11.9$ \\
2014 Dec 20.2 & 57011.2 & 121 & Swopes/CCD & $r$ & $>20.8$ & $>-12.1$ \\
2014 Dec 20.2 & 57011.2 & 121 & Swopes/CCD & $i$ & $>20.4$ & $>-12.5$ \\
2015 Jan 13.2 & 57035.2 & 145 & LCOGT-1m/CCD & $r$ & $>19.6$ & $>-13.3$ \\
2015 Jan 13.2 & 57035.2 & 145 & LCOGT-1m/CCD & $i$ & $>19.2$ & $>-13.7$ \\
2015 Jan 16.3 & 57038.3 & 148 & Lemmon-1.5m/2MASSCam & $J$ & $>13.5$ & $>-19.3$ \\
2015 Jan 16.3 & 57038.3 & 148 & Lemmon-1.5m/2MASSCam & $H$ & $>12.7$ & $>-20.1$ \\
2015 Jan 16.3 & 57038.3 & 148 & Lemmon-1.5m/2MASSCam & $K_s$ & $>12.6$ & $>-20.2$ \\
2015 Jan 20.0 & 57042.0 & 152 & Baade/IMACS & $\text{WB6226-7171}$ & $21.4$ $(0.1)$ & $-11.5$ \\
2015 Jan 29.1 & 57051.1 & 161 & du Pont/RetroCam & $Y$ & $19.9$ $(0.2)$ & $-12.9$ \\
2015 Jan 29.1 & 57051.1 & 161 & du Pont/RetroCam & $J$ & $19.8$ $(0.1)$ & $-13.0$ \\
2015 Jan 29.1 & 57051.1 & 161 & du Pont/RetroCam & $H$ & $18.8$ $(0.1)$ & $-14.0$ \\
2015 Feb 02.2 & 57055.2 & 165 & Swopes/CCD & $g$ & $>20.2$ & $>-12.8$ \\
2015 Feb 02.2 & 57055.2 & 165 & Swopes/CCD & $r$ & $>20.1$ & $>-12.8$ \\
2015 Feb 02.2 & 57055.2 & 165 & Swopes/CCD & $i$ & $>20.1$ & $>-12.8$ \\
2015 Feb 04.4 & 57057.4 & 167 & \textit{Spitzer}/IRAC & $[3.6]$ & $18.7$ $(0.2)$ & $-14.1$ \\
2015 Feb 04.4 & 57057.4 & 167 & \textit{Spitzer}/IRAC & $[4.5]$ & $15.68$ $(0.02)$ & $-17.1$ \\
2015 Feb 13.0 & 57066.0 & 176 & LCOGT-1m/CCD & $r$ & $>19.0$ & $>-13.9$ \\
2015 Feb 13.0 & 57066.0 & 176 & LCOGT-1m/CCD & $i$ & $>18.9$ & $>-14.0$ \\
2015 Feb 18.0 & 57071.0 & 181 & Baade/FourStar & $K_s$ & $19.17$ $(0.08)$ & $-13.6$ \\
2015 Feb 20.2 & 57073.2 & 183 & Lemmon-1.5m/2MASSCam & $J$ & $>14.3$ & $>-18.5$ \\
2015 Feb 20.2 & 57073.2 & 183 & Lemmon-1.5m/2MASSCam & $H$ & $>13.5$ & $>-19.3$ \\
2015 Feb 20.2 & 57073.2 & 183 & Lemmon-1.5m/2MASSCam & $K_s$ & $>13.2$ & $>-19.6$ \\
2015 Mar 09.1 & 57090.1 & 200 & du Pont/RetroCam & $Y$ & $>20.3$ & $>-12.6$ \\
2015 Mar 09.1 & 57090.1 & 200 & du Pont/RetroCam & $J$ & $>19.6$ & $>-13.2$ \\
2015 Mar 09.1 & 57090.1 & 200 & du Pont/RetroCam & $H$ & $>18.8$ & $>-14.0$ \\
2015 Mar 13.1 & 57094.1 & 204 & Swopes/CCD & $g$ & $>20.6$ & $>-12.4$ \\
2015 Mar 13.1 & 57094.1 & 204 & Swopes/CCD & $r$ & $>19.9$ & $>-13.0$ \\
2015 Mar 13.1 & 57094.1 & 204 & Swopes/CCD & $i$ & $>19.5$ & $>-13.4$ \\
2015 Mar 14.0 & 57095.0 & 205 & du Pont/RetroCam & $Y$ & $>20.2$ & $>-12.6$ \\
2015 Mar 14.0 & 57095.0 & 205 & du Pont/RetroCam & $J$ & $>20.3$ & $>-12.5$ \\
2015 Mar 14.0 & 57095.0 & 205 & du Pont/RetroCam & $H$ & $>19.4$ & $>-13.4$ \\
2015 Mar 15.0 & 57096.0 & 206 & LCOGT-1m/CCD & $r$ & $>19.2$ & $>-13.7$ \\
2015 Mar 15.0 & 57096.0 & 206 & LCOGT-1m/CCD & $i$ & $>18.8$ & $>-14.1$ \\
2015 Mar 15.0 & 57096.0 & 206 & Kuiper/Mont4k & $R$ & $>20.5$ & $>-12.4$ \\
2015 Apr 05.1 & 57117.1 & 227 & du Pont/RetroCam & $Y$ & $>20.2$ & $>-12.7$ \\
2015 Apr 05.1 & 57117.1 & 227 & du Pont/RetroCam & $J$ & $>19.3$ & $>-13.5$ \\
2015 Apr 05.1 & 57117.1 & 227 & du Pont/RetroCam & $H$ & $>18.5$ & $>-14.3$ \\
2015 Apr 16.4 & 57128.4 & 238 & LCOGT-1m/CCD & $r$ & $>20.1$ & $>-12.8$ \\
2015 Apr 16.4 & 57128.4 & 238 & LCOGT-1m/CCD & $i$ & $>19.5$ & $>-13.4$ \\
2015 Apr 30.0 & 57142.0 & 252 & du Pont/RetroCam & $Y$ & $>20.6$ & $>-12.3$ \\
2015 Apr 30.0 & 57142.0 & 252 & du Pont/RetroCam & $J$ & $>20.1$ & $>-12.7$ \\
2015 Apr 30.0 & 57142.0 & 252 & du Pont/RetroCam & $H$ & $>18.9$ & $>-13.8$ \\
2015 May 26.9 & 57168.9 & 278 & \textit{Spitzer}/IRAC & $[3.6]$ & $>18.8$ & $>-14.0$ \\
2015 May 26.9 & 57168.9 & 278 & \textit{Spitzer}/IRAC & $[4.5]$ & $17.01$ $(0.06)$ & $-15.7$ \\
2015 Jun 03.8 & 57176.8 & 286 & \textit{Spitzer}/IRAC & $[3.6]$ & $>18.9$ & $>-13.9$ \\
2015 Jun 03.8 & 57176.8 & 286 & \textit{Spitzer}/IRAC & $[4.5]$ & $17.05$ $(0.07)$ & $-15.7$ \\
2015 Jun 24.1 & 57197.1 & 307 & \textit{Spitzer}/IRAC & $[3.6]$ & $>18.7$ & $>-14.1$ \\
2015 Jun 24.1 & 57197.1 & 307 & \textit{Spitzer}/IRAC & $[4.5]$ & $17.32$ $(0.08)$ & $-15.4$ \\
2015 Sep 05.4 & 57270.4 & 380 & Baade/FourStar & $J$ & $>20.2$ & $>-12.6$ \\
2015 Sep 05.4 & 57270.4 & 380 & Baade/FourStar & $H$ & $>19.9$ & $>-12.9$ \\
2015 Sep 05.4 & 57270.4 & 380 & Baade/FourStar & $K_s$ & $>19.5$ & $>-13.3$ \\
2015 Nov 19.3 & 57345.3 & 455 & du Pont/RetroCam & $Y$ & $>20.9$ & $>-11.9$ \\
2015 Nov 19.3 & 57345.3 & 455 & du Pont/RetroCam & $J$ & $>20.7$ & $>-12.1$ \\
2015 Nov 19.3 & 57345.3 & 455 & du Pont/RetroCam & $H$ & $>19.8$ & $>-13.0$ \\
2015 Nov 22.3 & 57348.3 & 458 & du Pont/RetroCam & $Y$ & $>20.7$ & $>-12.2$ \\
2015 Nov 22.3 & 57348.3 & 458 & du Pont/RetroCam & $J$ & $>20.1$ & $>-12.7$ \\
2015 Nov 22.3 & 57348.3 & 458 & du Pont/RetroCam & $H$ & $>19.5$ & $>-13.3$ \\
2015 Dec 23.0 & 57379.0 & 489 & \textit{Spitzer}/IRAC & $[3.6]$ & $>18.8$ & $>-14.0$ \\
2015 Dec 23.0 & 57379.0 & 489 & \textit{Spitzer}/IRAC & $[4.5]$ & $>18.5$ & $>-14.3$ \\
2015 Dec 30.1 & 57386.1 & 496 & \textit{Spitzer}/IRAC & $[3.6]$ & $>18.7$ & $>-14.0$ \\
2015 Dec 30.1 & 57386.1 & 496 & \textit{Spitzer}/IRAC & $[4.5]$ & $>18.4$ & $>-14.3$ \\
2016 Jan 12.1 & 57399.1 & 509 & \textit{Spitzer}/IRAC & $[3.6]$ & $>18.8$ & $>-14.0$ \\
2016 Jan 12.1 & 57399.1 & 509 & \textit{Spitzer}/IRAC & $[4.5]$ & $>18.3$ & $>-14.4$
\enddata
\tablenotetext{a}{Phase is number of days since the earliest detection of this event on 2014 August 21.4 ($\mathrm{MJD} = 56890.4$).}
\tablenotetext{b}{1-$\sigma$ uncertainties are given in parentheses.}
\tablenotetext{c}{5-$\sigma$ limiting magnitudes are given for non-detections.}
\tablenotetext{d}{Absolute magnitudes corrected for Galactic extinction for IC~2163 from NED.}
\end{deluxetable*}

\begin{deluxetable*}{lcccccc}
\tablecaption{Photometry of SPIRITS\,14buu \label{table:phot_14buu}}
\tablehead{\colhead{UT Date} & \colhead{MJD} & \colhead{Phase\tablenotemark{a}} & \colhead{Tel./Inst.} & \colhead{Band} & \colhead{Apparent Magnitude\tablenotemark{b,c}} & \colhead{Absolute Magnitude\tablenotemark{c,d}} \\ 
\colhead{} & \colhead{} & \colhead{(days)} & \colhead{} & \colhead{} & \colhead{(mag)} & \colhead{(mag)} } 
\startdata
2013 Dec 16.1 & 56642.1 & -8 & NOTCam & $K_s$ & $>16.6$ & $>-16.2$ \\
2013 Dec 24.3 & 56650.3 & 0 & P60 & $g$ & $>20.4$ & $>-12.6$ \\
2013 Dec 24.3 & 56650.3 & 0 & P60 & $r$ & $20.0$ $(0.3)$ & $-13.0$ \\
2013 Dec 24.3 & 56650.3 & 0 & P60 & $i$ & $19.4$ $(0.3)$ & $-13.5$ \\
2014 Jan 10.2 & 56667.2 & 17 & du Pont/RetroCam & $Y$ & $18.20$ $(0.06)$ & $-14.6$ \\
2014 Jan 10.2 & 56667.2 & 17 & du Pont/RetroCam & $J$ & $17.80$ $(0.08)$ & $-15.0$ \\
2014 Jan 10.2 & 56667.2 & 17 & du Pont/RetroCam & $H$ & $17.44$ $(0.07)$ & $-15.3$ \\
2014 Jan 13.9 & 56670.9 & 21 & \textit{Spitzer}/IRAC & $[3.6]$ & $16.69$ $(0.08)$ & $-16.1$ \\
2014 Jan 13.9 & 56670.9 & 21 & \textit{Spitzer}/IRAC & $[4.5]$ & $16.70$ $(0.05)$ & $-16.1$ \\
2014 Feb 18.2 & 56706.2 & 56 & Swopes/CCD & $g$ & $>20.9$ & $>-12.2$ \\
2014 Feb 18.2 & 56706.2 & 56 & Swopes/CCD & $r$ & $20.0$ $(0.2)$ & $-12.9$ \\
2014 Feb 18.2 & 56706.2 & 56 & Swopes/CCD & $i$ & $19.2$ $(0.1)$ & $-13.7$ \\
2014 Mar 07.1 & 56723.1 & 73 & Baade/FourStar & $J$ & $18.09$ $(0.02)$ & $-14.7$ \\
2014 Mar 14.0 & 56730.0 & 80 & Swopes/CCD & $g$ & $>20.8$ & $>-12.2$ \\
2014 Mar 14.0 & 56730.0 & 80 & Swopes/CCD & $r$ & $20.0$ $(0.2)$ & $-13.0$ \\
2014 Mar 14.0 & 56730.0 & 80 & Swopes/CCD & $i$ & $19.4$ $(0.1)$ & $-13.5$ \\
2014 Mar 15.1 & 56731.1 & 81 & du Pont/RetroCam & $Y$ & $18.43$ $(0.08)$ & $-14.4$ \\
2014 Mar 15.1 & 56731.1 & 81 & du Pont/RetroCam & $J$ & $18.11$ $(0.06)$ & $-14.7$ \\
2014 Mar 16.1 & 56732.1 & 82 & du Pont/RetroCam & $H$ & $17.6$ $(0.1)$ & $-15.2$ \\
2014 Apr 13.0 & 56760.0 & 110 & Swopes/CCD & $g$ & $>20.8$ & $>-12.2$ \\
2014 Apr 13.0 & 56760.0 & 110 & Swopes/CCD & $r$ & $>20.6$ & $>-12.4$ \\
2014 Apr 13.0 & 56760.0 & 110 & Swopes/CCD & $i$ & $>20.5$ & $>-12.4$ \\
2014 May 13.7 & 56790.7 & 140 & \textit{Spitzer}/IRAC & $[3.6]$ & $18.0$ $(0.2)$ & $-14.8$ \\
2014 May 13.7 & 56790.7 & 140 & \textit{Spitzer}/IRAC & $[4.5]$ & $17.13$ $(0.06)$ & $-15.6$ \\
2014 Jun 08.7 & 56816.7 & 166 & \textit{Spitzer}/IRAC & $[3.6]$ & $18.4$ $(0.2)$ & $-14.3$ \\
2014 Jun 08.7 & 56816.7 & 166 & \textit{Spitzer}/IRAC & $[4.5]$ & $17.28$ $(0.08)$ & $-15.5$ \\
2014 Sep 29.4 & 56929.4 & 279 & Swopes/CCD & $g$ & $>21.5$ & $>-11.5$ \\
2014 Sep 29.4 & 56929.4 & 279 & Swopes/CCD & $r$ & $>21.0$ & $>-11.9$ \\
2014 Sep 29.4 & 56929.4 & 279 & Swopes/CCD & $i$ & $>20.7$ & $>-12.2$ \\
2014 Dec 05.3 & 56996.3 & 346 & du Pont/RetroCam & $Y$ & $>20.5$ & $>-12.3$ \\
2014 Dec 05.3 & 56996.3 & 346 & du Pont/RetroCam & $H$ & $>20.1$ & $>-12.7$ \\
2014 Dec 20.2 & 57011.2 & 361 & Swopes/CCD & $g$ & $>21.4$ & $>-11.6$ \\
2014 Dec 20.2 & 57011.2 & 361 & Swopes/CCD & $r$ & $>21.0$ & $>-11.9$ \\
2014 Dec 20.2 & 57011.2 & 361 & Swopes/CCD & $i$ & $>20.6$ & $>-12.3$ \\
2015 Jan 29.1 & 57051.1 & 401 & du Pont/RetroCam & $Y$ & $>21.2$ & $>-11.6$ \\
2015 Jan 29.1 & 57051.1 & 401 & du Pont/RetroCam & $J$ & $>20.6$ & $>-12.2$ \\
2015 Jan 29.1 & 57051.1 & 401 & du Pont/RetroCam & $H$ & $>20.2$ & $>-12.6$ \\
2015 Feb 02.2 & 57055.2 & 405 & Swopes/CCD & $g$ & $>20.1$ & $>-12.9$ \\
2015 Feb 02.2 & 57055.2 & 405 & Swopes/CCD & $r$ & $>20.1$ & $>-12.8$ \\
2015 Feb 02.2 & 57055.2 & 405 & Swopes/CCD & $i$ & $>20.1$ & $>-12.8$ \\
2015 Feb 04.4 & 57057.4 & 407 & \textit{Spitzer}/IRAC & $[3.6]$ & $>18.6$ & $>-14.2$ \\
2015 Feb 04.4 & 57057.4 & 407 & \textit{Spitzer}/IRAC & $[4.5]$ & $>17.8$ & $>-15.0$ \\
2015 Feb 18.0 & 57071.0 & 421 & Baade/FourStar & $K_s$ & $>19.1$ & $>-13.6$ \\
2015 Mar 09.1 & 57090.1 & 440 & du Pont/RetroCam & $Y$ & $>20.2$ & $>-12.6$ \\
2015 Mar 09.1 & 57090.1 & 440 & du Pont/RetroCam & $J$ & $>19.7$ & $>-13.1$ \\
2015 Mar 09.1 & 57090.1 & 440 & du Pont/RetroCam & $H$ & $>19.2$ & $>-13.6$ \\
2015 Mar 13.1 & 57094.1 & 444 & Swopes/CCD & $g$ & $>21.5$ & $>-11.5$ \\
2015 Mar 13.1 & 57094.1 & 444 & Swopes/CCD & $r$ & $>21.0$ & $>-12.0$ \\
2015 Mar 13.1 & 57094.1 & 444 & Swopes/CCD & $i$ & $>20.4$ & $>-12.5$ \\
2015 Mar 14.0 & 57095.0 & 445 & du Pont/RetroCam & $J$ & $>19.9$ & $>-12.9$ \\
2015 Apr 05.1 & 57117.1 & 467 & du Pont/RetroCam & $Y$ & $>20.0$ & $>-12.8$ \\
2015 Apr 05.1 & 57117.1 & 467 & du Pont/RetroCam & $J$ & $>19.4$ & $>-13.5$ \\
2015 Apr 05.1 & 57117.1 & 467 & du Pont/RetroCam & $H$ & $>18.9$ & $>-13.9$ \\
2015 Apr 30.0 & 57142.0 & 492 & du Pont/RetroCam & $Y$ & $>20.6$ & $>-12.2$ \\
2015 Apr 30.0 & 57142.0 & 492 & du Pont/RetroCam & $J$ & $>20.1$ & $>-12.7$ \\
2015 Apr 30.0 & 57142.0 & 492 & du Pont/RetroCam & $H$ & $>19.3$ & $>-13.5$ \\
2015 May 26.9 & 57168.9 & 519 & \textit{Spitzer}/IRAC & $[3.6]$ & $>18.8$ & $>-14.0$ \\
2015 May 26.9 & 57168.9 & 519 & \textit{Spitzer}/IRAC & $[4.5]$ & $>17.7$ & $>-15.0$ \\
2015 Jun 03.8 & 57176.8 & 526 & \textit{Spitzer}/IRAC & $[3.6]$ & $>18.7$ & $>-14.1$ \\
2015 Jun 03.8 & 57176.8 & 526 & \textit{Spitzer}/IRAC & $[4.5]$ & $>17.6$ & $>-15.1$ \\
2015 Jun 24.1 & 57197.1 & 547 & \textit{Spitzer}/IRAC & $[3.6]$ & $>18.4$ & $>-14.3$ \\
2015 Jun 24.1 & 57197.1 & 547 & \textit{Spitzer}/IRAC & $[4.5]$ & $>17.7$ & $>-15.1$ \\
2015 Sep 05.4 & 57270.4 & 620 & Baade/FourStar & $J$ & $>19.6$ & $>-13.2$ \\
2015 Sep 05.4 & 57270.4 & 620 & Baade/FourStar & $H$ & $>19.4$ & $>-13.4$ \\
2015 Sep 05.4 & 57270.4 & 620 & Baade/FourStar & $K_s$ & $>19.0$ & $>-13.8$ \\
2015 Nov 22.3 & 57348.3 & 698 & du Pont/RetroCam & $Y$ & $>20.8$ & $>-12.0$ \\
2015 Nov 22.3 & 57348.3 & 698 & du Pont/RetroCam & $H$ & $>20.0$ & $>-12.8$ \\
2015 Dec 23.0 & 57379.0 & 729 & \textit{Spitzer}/IRAC & $[3.6]$ & $>18.5$ & $>-14.3$ \\
2015 Dec 23.0 & 57379.0 & 729 & \textit{Spitzer}/IRAC & $[4.5]$ & $>17.6$ & $>-15.1$ \\
2015 Dec 30.1 & 57386.1 & 736 & \textit{Spitzer}/IRAC & $[3.6]$ & $>18.4$ & $>-14.4$ \\
2015 Dec 30.1 & 57386.1 & 736 & \textit{Spitzer}/IRAC & $[4.5]$ & $>17.8$ & $>-15.0$ \\
2016 Jan 12.1 & 57399.1 & 749 & \textit{Spitzer}/IRAC & $[3.6]$ & $>18.3$ & $>-14.5$ \\
2016 Jan 12.1 & 57399.1 & 749 & \textit{Spitzer}/IRAC & $[4.5]$ & $>17.7$ & $>-15.1$ \\
2016 Mar 02.1 & 57449.1 & 799 & du Pont/RetroCam & $Y$ & $>20.4$ & $>-12.5$ \\
2016 Mar 02.1 & 57449.1 & 799 & du Pont/RetroCam & $H$ & $>18.5$ & $>-14.2$
\enddata
\tablenotetext{a}{Phase is number of days since the earliest detection of this event on 2014 August 21.4 ($\mathrm{MJD} = 56650.3$).}
\tablenotetext{b}{1-$\sigma$ uncertainties are given in parentheses.}
\tablenotetext{c}{5-$\sigma$ limiting magnitudes are given for non-detections.}
\tablenotetext{d}{Absolute magnitudes corrected for Galactic extinction for IC~2163 from NED.}
\end{deluxetable*}

\begin{deluxetable}{cccc}
\tablecaption{Parameters of SED Component Blackbody Fits\label{table:bb_fits}}
\tablehead{\colhead{Phase\tablenotemark{a}} & Total Extinction\tablenotemark{b}, $A_V$ & $T_{\mathrm{eff}}$ & $R_{\mathrm{BB}}$ \\ 
\colhead{(days)} & \colhead{(mag)} & \colhead{(K)} & \colhead{(cm)}} 
\startdata
40   & 0.238 & 3300 & $1.4\times10^{15}$ \\
     & 1.0   & 3800 & $1.3\times10^{15}$ \\
		 & 2.0   & 4600 & $1.1\times10^{15}$ \\
		 & 2.2   & 4900 & $1.1\times10^{15}$ \\
 		 & 3.0   & 5900 & $9.6\times10^{14}$ \\
\hline
\rule{0pt}{2.6ex} 
63   & 0.238 & 4000 & $6.9\times10^{14}$ \\
     & 1.0   & 4700 & $6.1\times10^{14}$ \\
		 & 2.0   & 6100 & $5.1\times10^{14}$ \\
		 & 2.2   & 6700 & $4.9\times10^{14}$ \\
		 & 3.0   &  9300 & $4.0\times10^{14}$ \\
\hline
\rule{0pt}{2.6ex} 
106  & 0.238 & 3500 & $6.3\times10^{14}$ \\
     & 1.0   & 3700 & $6.2\times10^{14}$ \\
		 & 2.0   & 4000 & $6.1\times10^{14}$ \\
		 & 2.2   & 4100 & $6.1\times10^{14}$ \\
		 & 3.0   & 4400 &  $5.9\times10^{14}$ \\
\hline
\rule{0pt}{2.6ex} 
161  & 0.238 & 2700 & $5.7\times10^{14}$ \\
     & 1.0   & 2800 & $5.5\times10^{14}$ \\
		 & 2.0   & 3100 & $5.3\times10^{14}$ \\
		 & 2.2   & 3200 & $5.2\times10^{14}$ \\
		 & 3.0   & 3400 & $5.1\times10^{14}$ \\
\hline
\rule{0pt}{2.6ex}
167   & 0.238 & $270$ & $3.7\times10^{17}$ \\
      & 1.0   & $270$ & $3.7\times10^{17}$ \\
		  & 2.0   & $270$ & $3.6\times10^{17}$ \\
		  & 2.24   & $270$ & $3.6\times10^{17}$ \\
		  & 3.0   & $270$ & $3.5\times10^{17}$
\enddata
\tablenotetext{a}{Phase is number of days since the earliest detection of this event on 2014 August 21.4 ($\mathrm{MJD} = 56890.4$).}
\tablenotetext{b}{Assumed total extinction to SPIRITS\,15c including both Galactic ($A_V = 0.238$), and any additional extinction, e.g., from the host environment.}
\end{deluxetable}

\begin{deluxetable}{cc}
\tablecaption{Spectral line identifications\tablenotemark{a,b}}
\tablehead{Species & Rest Wavelength (air) \\ 
                   & $(\mu\mathrm{m})$}
\startdata
He~\textsc{i}    & 1.083 \\
                 & 2.058 \\
\hline
\rule{0pt}{2.6ex}
C~\textsc{i}     & 1.175 \\
\hline
\rule{0pt}{2.6ex}
O~\textsc{i}     & 0.926 \\
                 & 1.129 \\
							   & 1.130 \\
                 & 1.317 \\
\hline
\rule{0pt}{2.6ex}
Na~\textsc{i}    & 1.140  \\
\hline
\rule{0pt}{2.6ex}								 
Mg~\textsc{i}    & 1.504 \\
\hline
\rule{0pt}{2.6ex}
Si~\textsc{i}    & 1.203  \\
\hline
\rule{0pt}{2.6ex}
$\mathrm{[Si~\textsc{i}]}$  & 1.083  \\
                 & 1.098  \\
								 & 1.129  \\
								 & 1.646  \\
\hline
\rule{0pt}{2.6ex}
$\mathrm{Ca~\textsc{ii}}$ & 0.845 \\ 
                & 0.854 \\
								& 0.866 \\
\hline
\rule{0pt}{2.6ex}								
$\mathrm{[Co~\textsc{ii}]}$ & 0.934 \\
                 & 0.934 \\
								 & 1.019 \\
								 & 1.025 \\
								 & 1.028 \\
								 & 1.547  \\
\hline
\rule{0pt}{2.6ex}
$\mathrm{[Fe~\textsc{ii}]}$ & 1.257  \\
                 & 1.321  \\
								 & 1.328  \\
								 & 1.644  \\
\hline
\rule{0pt}{2.6ex}							 
$^{12}$C$^{16}$O & 2.294 \\
                 & 2.323 \\
								 & 2.354 \\
								 & 2.383 \\
								 & 2.414 \\
								 & 2.446
\enddata
\tablenotetext{a}{Line identifications come from \citet{ergon14}, \citet{ergon15}, and \citet{jerkstrand15}.}
\tablenotetext{b}{Wavelengths listed for $^{12}$C$^{16}$O correspond to band heads of the $\Delta v = 2$ vibrational overtones of this molecule.}
\end{deluxetable}

\twocolumngrid

\end{document}